\documentclass[a4paper,11pt]{article}
\pdfoutput=1 % if your are submitting a pdflatex (i.e. if you have
\usepackage{jheppub} % for details on the use of the package, please
\makeatletter
\DeclareRobustCommand*{\bfseries}{%
  \not@math@alphabet\bfseries\mathbf
  \fontseries\bfdefault\selectfont
  \boldmath
}
\makeatother

\usepackage{calc}
\usepackage{cleveref}
\usepackage{rotating}
\usepackage[english]{babel}
\usepackage{graphicx}
\usepackage{subfig}
\usepackage{float}
\usepackage{amsmath}
\usepackage{amssymb}
\usepackage{amsthm}
\usepackage{latexsym}
\usepackage{dcolumn}
\newcommand{\newc}{\newcommand*}

\long\def\begincomment#1\endcomment{%
        \begingroup\sf\baselineskip12pt#1\endgroup}

\newc{\etal}{\textrm{et al.}} 
\newc{\eg}{\textrm{e.g.}} 
\newc{\ie}{\textrm{i.e.}}
\newc{\etc}{\textrm{etc}}
\newc\vs{\textrm{vs.}}
\newc{\cl}{\rm {C.L.}}
\newc{\ev}{\ensuremath{\,\mathrm{eV}}}
\newc{\kev}{\ensuremath{\,\mathrm{keV}}}
\newc{\mev}{\ensuremath{\,\mathrm{MeV}}}
\newc{\gev}{\ensuremath{\,\mathrm{GeV}}}
\newc{\tev}{\ensuremath{\,\mathrm{TeV}}}
\newc{\MeV}{\mev} 
\newc{\TeV}{\tev}
\newc{\invpb}{\ensuremath{/\text{pb}}}
\newc{\invfb}{\ensuremath{\,\text{fb}^{-1}}}
\newc\nb{\ensuremath{\,\mathrm{nb}}} \newc\pb{\ensuremath{\,\mathrm{pb}}} \newc\fb{\ensuremath{\,\mathrm{fb}}}
\newc\pc{\ensuremath{\,\mathrm{pc}}}
\newc\kpc{\ensuremath{\,\mathrm{kpc}}}
\newc\mpc{\ensuremath{\,\mathrm{Mpc}}}
\newc\ps{\ensuremath{\,\mathrm{ps}}} 
\newc\cmeter{\ensuremath{\,\mathrm{cm}}} 
\newc\meter{\ensuremath{\,\mathrm{m}}} 
\newc\kmeter{\ensuremath{\,\mathrm{km}}}
\newc\second{\ensuremath{\,\mathrm{s}}}
\newc\msecond{\ensuremath{\,\mathrm{ms}}}
\newc\nsecond{\ensuremath{\,\mathrm{ns}}}
\newc\psecond{\ensuremath{\,\mathrm{ps}}}
\newc{\chisqmin}{\ensuremath{\chi^2_{\mathrm{min}}}}
\newc{\Delchisq}{\ensuremath{\Delta\chi^2}}
\newc{\chisq}{\ensuremath{\chi^2}}
\newc{\like}{\ensuremath{\mathcal{L}}}
\newc\lsim{\ensuremath{\mathrel{\rlap{\lower4pt\hbox{\hskip1pt$\sim$}}\raise1pt\hbox{$<$}}}}
\newc\gsim{\ensuremath{\mathrel{\rlap{\lower4pt\hbox{\hskip1pt$\sim$}}\raise1pt\hbox{$>$}}}}
\newc{\VEV}[1]{\ensuremath{\langle #1 \rangle}}
\newc{\dl}{\ensuremath{\stackrel{\leftarrow}{D}}}
\newc{\dr}{\ensuremath{\stackrel{\rightarrow}{D}}}
\newc{\bcenter}{\begin{center}}    \newc{\ecenter}{\end{center}}
\newc{\bfl}{\begin{flushleft}}    \newc{\efl}{\end{flushleft}}
\newc{\bfr}{\begin{flushright}}    \newc{\efr}{\end{flushright}}

\newc{\bi}{\begin{itemize}}
\newc{\ei}{\end{itemize}}
\newc{\bed}{\begin{description}}
\newc{\eed}{\end{description}}
\newc{\ben}{\begin{enumerate}}
\newc{\een}{\end{enumerate}}

\newc{\be}{\begin{equation}}
\newc{\ee}{\end{equation}}
\newc{\bea}{\begin{eqnarray}}
\newc{\eea}{\end{eqnarray}}
\newc{\ra}{\rightarrow}
\newc{\alphas}{\ensuremath{\alpha_s}}
\newc{\alphatwo}{\ensuremath{\alpha_2}}
\newc{\alphaone}{\ensuremath{\alpha_1}}
\newc{\alphai}[1]{\ensuremath{\alpha_{#1}}}
\newc{\alphaem}{\ensuremath{\alpha_{\mathrm{em}}}}
\newc{\alphaeff}{\ensuremath{\alpha_{\mathrm{eff}}}}
\newc{\sineff}{\ensuremath{\sin \theta_{\mathrm{eff}}}}
\newc{\sinsqeff}{\ensuremath{\sin^2 \theta_{\mathrm{eff}}}}
\newc{\dalphahad}{\ensuremath{\Delta \alpha_{\mathrm{had}}}}
\newc{\yt}{\ensuremath{h_t}} \newc{\yb}{\ensuremath{h_b}} \newc{\ytau}{\ensuremath{h_{\tau}}}
\newc\mz{\ensuremath{M_Z}} 
\newc\mw{\ensuremath{m_W}}
\newc\mZ{\mz}        \newc\mW{\mw}
\newc\mhsm{\ensuremath{ m_{H_{\mathrm{SM}}}}}
\newc{\mtop}{\ensuremath{ m_t}}               \newc{\mtpole}{\ensuremath{ M_t}}
\newc{\mbottom}{\ensuremath{ m_b}} 
\newc{\mtau}{\ensuremath{ m_{\tau}}}
\newc{\mt}{\mtpole}
\newc{\mb}{\mbottom} 
\newc{\rgg}{\ensuremath{R_{h}(\gamma\gamma)}}
\newc{\rzz}{\ensuremath{R_{h}(ZZ)}}
\newc{\rtwogg}{\ensuremath{R_{h_2}(\gamma\gamma)}}
\newc{\rtwozz}{\ensuremath{R_{h_2}(ZZ)}}
\newc{\ronegg}{\ensuremath{R_{h_1}(\gamma\gamma)}}
\newc{\ronezz}{\ensuremath{R_{h_1}(ZZ)}}
\newc{\rsiggg}{\ensuremath{R_{h_\textrm{sig}}(\gamma\gamma)}}
\newc{\rsigzz}{\ensuremath{R_{h_\textrm{sig}}(ZZ)}}
\newc{\llbar}{\ensuremath{\ell\bar{\ell}}}
\newc{\tauptaum}{\ensuremath{ \tau^+\tau^-}}
\newc{\qqbar}{\ensuremath{ q\bar{q}}} \newc{\ppbar}{\ensuremath{ p\bar{p}}}
\newc{\bbbar}{\ensuremath{ b\bar{b}}} \newc{\ttbar}{\ensuremath{ t\bar{t}}}
\newc{\ffbar}{\ensuremath{ f\bar{f}}} \newc{\tautaubar}{\ensuremath{ \tau\bar{\tau}}}
\newc{\mchi}{\ensuremath{m_{\chi}}}
\newc{\squark}{\ensuremath{\tilde{q}}}
\newc{\slepton}{\ensuremath{\tilde{l}}}
\newc{\gluino}{\ensuremath{\tilde{g}}} 
\newc{\wino}{\ensuremath{\tilde{W}}}
\newc{\bino}{\ensuremath{\tilde{B}}}
\newc{\mgluino}{\ensuremath{{m_{\gluino}}}}
\newc{\tone}{\ensuremath{{\tilde{t}_1}}}
\newc{\sthw}{\ensuremath{ \sin\theta_W}}              \newc{\cthw}{\ensuremath{\cos\theta_W}}
\newc{\tanthw}{\ensuremath{ \tan\theta_W}}              \newc{\cotthw}{\ensuremath{\cot\theta_W}}
\newc{\ssqthw}{\ensuremath{\sin^2 \theta_W}}
\newc{\msbar}{\ensuremath{\overline{MS}}} \newc{\drbar}{\ensuremath{\overline{DR}}}
\newc{\mtmtsmmsbar}{\ensuremath{ m_t(m_t)^{\msbar}_{{\mathrm{SM}}}}}
\newc{\mtmtsmdrbar}{\ensuremath{ m_t(m_t)^{\drbar}_{{\mathrm{SM}}}}}
\newc{\mtmtmssmdrbar}{\ensuremath{ m_t(m_t)^{\drbar}_{{\mathrm{SUSY}}}}}
\newc{\mbmbmsbar}{\ensuremath{ m_b(m_b)^{\msbar} }}
\newc{\mbmbsmmsbar}{\ensuremath{ m_b(m_b)^{\msbar}_{{\mathrm{SM}}}}}
\newc{\mbmzsmmsbar}{\ensuremath{ m_b(\mz)^{\msbar}_{{\mathrm{SM}}}}}
\newc{\mbmzsmdrbar}{\ensuremath{ m_b(\mz)^{\drbar}_{{\mathrm{SM}}}}}
\newc{\mbmzmssmdrbar}{\ensuremath{ m_b(\mz)^{\drbar}_{{\mathrm{SUSY}}}}}
\newc{\mtaumzsmmsbar}{\ensuremath{ m_{\tau}(\mz)^{\msbar}_{{\mathrm{SM}}}}}
\newc{\mtaumzsmdrbar}{\ensuremath{ m_{\tau}(\mz)^{\drbar}_{{\mathrm{SM}}}}}
\newc{\mtaumzmssmdrbar}{\ensuremath{ m_{\tau}(\mz)^{\drbar}_{{\mathrm{SUSY}}}}}
\newc{\alphasmzms}{\ensuremath{\alpha_s(M_Z)^{\overline{MS}}}}
\newc{\alphaimzms}[1]{\ensuremath{\alpha_{#1}(M_Z)^{\overline{MS}}}}
\newc{\alphaemmz}{\ensuremath{\alpha_{\mathrm{em}}(M_Z)^{\overline{MS}}}}
\newc{\mzero}{\ensuremath{{m_0}}}
\newc{\mhalf}{\ensuremath{ m_{1/2}}}
\newc{\tanb}{\ensuremath{\tan\beta}}
\newc{\azero}{\ensuremath{ A_0}}
\newc{\bzero}{\ensuremath{ B_0}}
\newc{\signmu}{\ensuremath{\rm{sgn}\,\mu}}
\newc{\mueff}{\ensuremath{\mu_{\rm{eff}}}}
\newc{\lam}{\ensuremath{{\lambda}}}
\newc{\kap}{\ensuremath{{\kappa}}}
\newc{\alam}{\ensuremath{{A_{\lambda}}}}
\newc{\akap}{\ensuremath{{A_{\kappa}}}}
\newc{\hs}{\ensuremath{ H_s}}      
\newc{\mhs}{\ensuremath{ m_{H_s}}} 
\newc{\mgut}{\ensuremath{ M_{\rm GUT}}}
\newc{\mplanck}{\ensuremath{ M_{\rm P}}}      \newc{\mpl}{\ensuremath{ M_{\rm Pl}}}
\newc{\msusy}{\ensuremath{ M_{\rm SUSY}}}      \newc{\ms}{\ensuremath{ M_{\rm S}}}
 \newc{\mhl}{\ensuremath{m_\hl}} 
 \newc{\mhone}{\ensuremath{m_{h_1}}} 
 \newc{\mhtwo}{\ensuremath{m_{h_2}}} 
 \newc{\mglu}{\ensuremath{m_{\tilde g}}} 
 \newc{\mul}{\ensuremath{m_{\tilde{u}_L}}} 
 \newc{\mtone}{\ensuremath{m_{\tilde{t}_1}}} 
 \newc{\ma}{\ensuremath{m_A}} 
 \newc{\maone}{\ensuremath{m_{a_1}}} 
 \newc{\matwo}{\ensuremath{m_{a_2}}}
 \newc{\hone}{\ensuremath{h_1}}
 \newc{\htwo}{\ensuremath{h_2}}
 \newc{\aone}{\ensuremath{a_1}}
 \newc{\atwo}{\ensuremath{a_2}}
 \newc{\mhu}{\ensuremath{ m_{H_u}}}       
 \newc{\mhd}{\ensuremath{ m_{H_d}}}
 \newc{\mhusq}{\ensuremath{ m_{H_u}^2}}       
 \newc{\mhdsq}{\ensuremath{ m_{H_d}^2}}
 \newc{\mhuew}{\ensuremath{ m^{\ast}_{H_u}}}       
 \newc{\mhdew}{\ensuremath{ m^{\ast}_{H_d}}}
 \newc{\mhuewsq}{\ensuremath{ m^{\ast\, 2}_{H_u}}}       
 \newc{\mhdewsq}{\ensuremath{ m^{\ast\, 2}_{H_d}}}
 \newc{\hu}{\ensuremath{ H_u}}       
 \newc{\hd}{\ensuremath{ H_d}}
 \newc{\barmhu}{\ensuremath{ \bar{m}_{H_u}}}
 \newc{\barmhd}{\ensuremath{ \bar{m}_{H_d}}}

 \newc{\mqthree}{\ensuremath{m_{\widetilde{Q}_3}^2}}
 \newc{\muthree}{\ensuremath{m_{\tilde{u}_3}^2}}
 \newc{\mdthree}{\ensuremath{m_{\tilde{d}_3}^2}}
 \newc{\mlthree}{\ensuremath{m_{\widetilde{L}_3}^2}}
 \newc{\methree}{\ensuremath{m_{\tilde{e}_3}^2}}
 \newc{\mqtwo}{\ensuremath{m_{\widetilde{Q}_2}^2}}
 \newc{\mutwo}{\ensuremath{m_{\tilde{u}_2}^2}}
 \newc{\mdtwo}{\ensuremath{m_{\tilde{d}_2}^2}}
 \newc{\mltwo}{\ensuremath{m_{\widetilde{L}_2}^2}}
 \newc{\metwo}{\ensuremath{m_{\tilde{e}_2}^2}}
 \newc{\mqone}{\ensuremath{m_{\widetilde{Q}_1}^2}}
 \newc{\muone}{\ensuremath{m_{\tilde{u}_1}^2}}
 \newc{\mdone}{\ensuremath{m_{\tilde{d}_1}^2}}
 \newc{\mlone}{\ensuremath{m_{\widetilde{L}_1}^2}}
 \newc{\meone}{\ensuremath{m_{\tilde{e}_1}^2}}
 \newc{\mone}{\ensuremath{M_1}}
 \newc{\monesq}{\ensuremath{M_1^2}}
 \newc{\mtwo}{\ensuremath{M_2}}
 \newc{\mtwosq}{\ensuremath{M_2^2}}
 \newc{\mthree}{\ensuremath{M_3}}
 \newc{\mthreesq}{\ensuremath{M_3^2}}
 \newc{\atau}{\ensuremath{{A_{\tau}}}}
 \newc{\at}{\ensuremath{{A_{t}}}}
 \newc{\ab}{\ensuremath{{A_{b}}}}
 \newc{\atausq}{\ensuremath{{A_{\tau}^2}}}
 \newc{\atsq}{\ensuremath{{A_{t}^2}}}
 \newc{\absq}{\ensuremath{{A_{b}^2}}}
 \newc{\dmzero}{\ensuremath{\Delta{_{m_0}}}}
 \newc{\dmhalf}{\ensuremath{\Delta{_{m_{1/2}}}}}
 \newc{\dmu}{\ensuremath{\Delta{_{\mu}}}}

 \newc{\pten}{\ensuremath{\psi_{10}}}
 \newc{\ffive}{\ensuremath{\phi_{5}}}
 \newc{\hfive}{\ensuremath{h_{5}}}
 \newc{\hbfive}{\ensuremath{h_{\bar{5}}}}
 \newc{\thet}{\ensuremath{\theta_{50}}}
 \newc{\thetb}{\ensuremath{\theta_{\,\overline{50}}}}
 \newc{\ptenhat}{\ensuremath{\hat{\psi}_{10}}}
 \newc{\ffivehat}{\ensuremath{\hat{\phi}_{5}}}
 \newc{\hfivehat}{\ensuremath{\hat{h}_{5}}}
 \newc{\hbfivehat}{\ensuremath{\hat{h}_{\bar{5}}}}
 \newc{\thethat}{\ensuremath{\hat{\theta}_{50}}}
 \newc{\thetbhat}{\ensuremath{\hat{\theta}_{\,\overline{50}}}}
 \newc{\si}{\ensuremath{\Sigma}}
 \newc{\mfive}{\ensuremath{m_5^2}}
 \newc{\mten}{\ensuremath{m_{10}^2}}
 \newc{\dfive}{\ensuremath{\Delta^2_5}}
 \newc{\dbfive}{\ensuremath{\Delta^2_{\bar{5}}}}
 \newc{\dfifty}{\ensuremath{\Delta^2_{50}}}
 \newc{\dfiftyb}{\ensuremath{\Delta^2_{\,\overline{50}}}}
 \newc{\msi}{\ensuremath{m_{\Sigma}^2}}
 \newc{\lamh}{\ensuremath{\lambda_{H}}}
 \newc{\lamhb}{\ensuremath{\lambda_{\bar{H}}}}
 \newc{\ah}{\ensuremath{A_{H}}}
 \newc{\ahb}{\ensuremath{A_{\bar{H}}}}
 \newc{\lams}{\ensuremath{\lambda_{S}}}
 \newc{\as}{\ensuremath{A_{S}}}
 \newc{\lamsig}{\ensuremath{\lambda_{\si}}}
 \newc{\asig}{\ensuremath{A_{\si}}}

 \newc{\msten}{\ensuremath{m_{16}^2}}
 \newc{\mhun}{\ensuremath{m_{126}^2}}
 \newc{\mhunb}{\ensuremath{m_{\bar{126}}^2}}
 \newc{\mthun}{\ensuremath{m_{210}^2}}
 \newc{\ahun}{\ensuremath{A_{\bar{126}}}}
 \newc{\yhun}{\ensuremath{Y_{\bar{126}}}}
 \newc{\aten}{\ensuremath{A_{10}}}
 \newc{\yten}{\ensuremath{Y_{10}}}
 \newc{\alone}{\ensuremath{A_{\lambda_1}}}
 \newc{\altwo}{\ensuremath{A_{\lambda_2}}}
 \newc{\althree}{\ensuremath{A_{\lambda_3}}}
 \newc{\althreeb}{\ensuremath{A_{\bar{\lambda_3}}}}
 \newc{\lone}{\ensuremath{\lambda_1}}
 \newc{\ltwo}{\ensuremath{\lambda_2}}
 \newc{\lthree}{\ensuremath{\lambda_3}}
 \newc{\lthreeb}{\ensuremath{\bar{\lambda_3}}}

\newc{\sigsip}{\ensuremath{\sigma^{\rm SI}_{p}}}	\newc{\sigsin}{\ensuremath{\sigma^{\rm SI}_{n}}}
\newc{\sigsdp}{\ensuremath{\sigma^{\rm SD}_{p}}}	\newc{\sigsdn}{\ensuremath{\sigma^{\rm SD}_{n}}}
\newc{\sigsi}{\ensuremath{\sigma^{\rm SI}}}	\newc{\sigsd}{\ensuremath{\sigma^{\rm SD}}}
\newc{\sigv}{\ensuremath{\sigma v}}
\newc{\abund}{\ensuremath{ \Omega h^2}}
\newc{\omegadm}{\ensuremath{ \Omega_{{\rm DM}}}}     \newc{\abunddm}{\ensuremath{ \Omega_{{\rm DM}} h^2}} 
\newc{\omegam}{\ensuremath{ \Omega_{{\rm m}}}}       \newc{\abundm}{\ensuremath{ \Omega_{{\rm m}} h^2}}
\newc{\omegab}{\ensuremath{ \Omega_{{\rm b}}}}	\newc{\abundb}{\ensuremath{ \Omega_{{\rm b}} h^2}}
\newc{\omegatot}{\ensuremath{ \Omega_{{\rm TOT}}}}
\newc{\omegacdm}{\ensuremath{ \Omega_{{\rm CDM}}}}   \newc{\abundcdm}{\ensuremath{ \Omega_{{\rm CDM}} h^2}}
\newc{\omegalambda}{\ensuremath{ \Omega_{\Lambda}}} \newc{\abundlambda}{\ensuremath{ \Omega_{\Lambda} h^2}}
\newc{\omegarad}{\ensuremath{ \Omega_{{\rm rad}}}}  \newc{\abundrad}{\ensuremath{ \Omega_{{\rm rad}} h^2}}
\newc{\rhocrit}{\ensuremath{ \rho_{\rm crit}}}
\newc{\rhochi}{\ensuremath{ \rho_{\chi}}}
\newc{\abunchi}{\ensuremath{\Omega_\chi h^2}}
\newc{\abundlsp}{\ensuremath{\Omega_{\rm LSP}h^2}}
\newc{\abundchi}{\ensuremath{\Omega_\chi h^2}}% For multiple citations
\newc{\tf}{\ensuremath{T_f}} \newc{\xf}{\ensuremath{x_f}}
\newc{\tr}{\ensuremath{T_R}}

\newc{\amu}{\ensuremath{ a_{\mu}}}        \newc{\amususy}{\ensuremath{ a_{\mu}^{\mathrm{SUSY}}}}
\newc{\amuexpt}{\ensuremath{ a_{\mu}^{\mathrm{expt}}}}        \newc{\amusm}{\ensuremath{ a_{\mu}^{\mathrm{SM}}}}
\newc\deltaamu{\ensuremath{\Delta a_{\mu}}} \newc{\deltaamususy}{\ensuremath{\delta a_{\mu}^{\mathrm{SUSY}}}}
\newc\gmtwo{\ensuremath{ (g-2)_{\mu}}} 
\newc{\deltagmtwomususy}{\ensuremath{\delta\left(g-2\right)_{\mu}^{\mathrm{SUSY}}}}
\newc{\deltagmtwomu}{\ensuremath{\delta\left(g-2\right)_{\mu}}}
\newc\BR{\ensuremath{\rm BR}}
\newc\bsgamma{\ensuremath{ b\rightarrow s \gamma }}
\newc\bxsgamma{\ensuremath{\overline{B}\rightarrow X_{s}\gamma}}
\newc\brbsgamma{\ensuremath{\BR\left(\bsgamma\right)}}
\newc\brbxsgamma{\ensuremath{\BR\left(\bxsgamma\right)}}
\newc\bsmumu{\ensuremath{B_s\to\mu^+\mu^-}}
\newc\brbsmumu{\ensuremath{\BR\left(B_s\to\mu^+\mu^-\right)}}
\newc\bdmmumu{\ensuremath{\overline{B}_d\to\mu^+\mu^-}}
\newc\bbbarmix{\ensuremath{\overline{B}_s\mbox{-}B_s}}      % B_s mixing
\newc\delmbs{\ensuremath{\Delta M_{B_s}}}
\newc{\butaunu}{\ensuremath{B_u \rightarrow \tau \nu}}
\newc{\brbutaunu}{\ensuremath{\BR\left(B_u \rightarrow \tau \nu\right)}}
\newcommand*{\reftable}[1]{Table~\ref{#1}}         
       
\newcommand*{\reffig}[1]{Fig.~\ref{#1}}
 
        \newcommand*{\refeq}[1]{Eq.~(\ref{#1})}
     \newcommand*{\refsec}[1]{Sec.~\ref{#1}}

\newcommand*{\charone}{\ensuremath{\chi^{\pm}_1}}

\newcommand*{\mstopone}{\ensuremath{m_{\tilde{t}_1}}}
\newcommand*{\mstoptwo}{\ensuremath{m_{\tilde{t}_2}}}

\let\oldcite\cite
\renewcommand*{\cite}{~\oldcite}

\newcommand*{\hl}{\ensuremath{h}}

\RequirePackage[normalem]{ulem}

\restylefloat{figure}

\title{Testing dark matter with Cherenkov light -- prospects of H.E.S.S. and CTA for exploring minimal supersymmetry}

\author[a]{Andrzej Hryczuk,}
\author[a]{Krzysztof Jod\l{}owski,}
\author[b]{Emmanuel Moulin,}
\author[b]{Lucia Rinchiuso,}
\author[c,a]{Leszek Roszkowski,}
\author[a]{Enrico Maria Sessolo}
\author[d,a]{and Sebastian Trojanowski}

\affiliation[a]{National Centre for Nuclear Research,\\
  Pasteura 7, 02-093 Warsaw, Poland}
\affiliation[b]{IRFU,CEA,D\'epartement de Physique des Particules, Universit\'e Paris-Saclay,\\F-91191 Gif-sur-Yvette, France}
\affiliation[c]{Astrocent, Nicolaus Copernicus Astronomical Center Polish Academy of Sciences,\\ul. Bartycka 18, 00-716 Warsaw, Poland}
\affiliation[d]{Consortium for Fundamental Physics, School of Mathematics and Statistics, University of Sheffield,\\Hounsfield Road, Sheffield, S3 7RH, UK}

\emailAdd{andrzej.hryczuk@ncbj.gov.pl}
\emailAdd{krzysztof.jodlowski@ncbj.gov.pl}
\emailAdd{emmanuel.moulin@cea.fr}
\emailAdd{lucia.rinchiuso@cea.fr}
\emailAdd{leszek.roszkowski@ncbj.gov.pl}
\emailAdd{enrico.sessolo@ncbj.gov.pl}
\emailAdd{s.trojanowski@sheffield.ac.uk}

\abstract{We provide an updated and improved study of the prospects of
  the H.E.S.S. and Cherenkov Telescope Array (CTA) experiments in
  testing neutralino dark matter in the Minimal Supersymmetric
  Standard Model with nine free parameters (p9MSSM). We include all
  relevant experimental constraints and theoretical developments, in
  particular a calculation of the Sommerfeld enhancement for both
  present-day annihilations and the relic abundance. We perform a
  state-of-the-art analysis of the CTA sensitivity with a
  log-likelihood test ratio statistics and apply it to a numerical
  scan of the p9MSSM parameter space focusing on a TeV scale dark
  matter. We find that, assuming Einasto profile of dark matter halo
  in the Milky Way, H.E.S.S. has already been able to nearly reach the
  so-called thermal WIMP value, while CTA will go below it by
  providing a further improvement of at least an order of
  magnitude. Both H.E.S.S. and CTA are sensitive to several cases for
  which direct detection cross section will be below the so-called
  neutrino floor, with H.E.S.S. being sensitive to most of the wino
  region, while CTA also covering a large fraction of the $\sim1\tev$
  higgsino region. We show that CTA sensitivity will be further
  improved in the monochromatic photon search mode for both
  single-component and underabundant dark matter.  }

\begin{document}
\maketitle
\flushbottom
%%%%%%%%%%%%%%%%%%%%%%%%%%%%%%%%%%%%%%%%%%%%%%%%%%%%%%%%%%%%%%%%
\section{\label{sec:intro}Introduction}
%%%%%%%%%%%%%%%%%%%%%%%%%%%%%%%%%%%%%%%%%%%%%%%%%%%%%%%%%%%%%&&&&

Dark matter (DM) is the dominant component of matter in the universe but its nature remains unknown. Dark matter in the form of weakly interacting massive particles (WIMPs) has attracted a great deal of interest in the last decades, and a worldwide experimental effort is underway to unveil its fundamental properties (for recent reviews see, \textit{e.g.},\cite{Arcadi:2017kky,Roszkowski:2017nbc}).
WIMP candidates appear in many extensions of the Standard Model (SM), among which a notable example is supersymmetry (SUSY). A multi-faceted approach has been developed to search for WIMP DM that exploits the complementarity of direct detection strategies, in which one attempts to detect WIMPs scattering off the target nuclei, indirect detection, which seeks detecting products of WIMP self annihilations, and the production at high-energy colliders.

The so-far null experimental searches carried out at colliders and in underground laboratories for the direct detection of DM have pushed the WIMP mass scale into the TeV range. In the SUSY framework, this is in agreement with expectations for the scale of soft SUSY-breaking consistent with the discovery at the LHC of a 125\gev\ Higgs boson. To be able to study TeV WIMPs at colliders requires center-of-mass energy beyond that of the Large Hadron Collider (LHC), and direct detection faces the lowered number density of DM particles due to the larger DM mass.

It is in this mass regime where indirect detection with gamma rays may
also play a major role. While a continuous part of the gamma-ray flux
is expected to drop for energies close to the DM mass, pronounced
line-like features that appear there provide a distinctive signature
of TeV DM over astrophysical backgrounds. The quest for such spectral
features further motivates the searches carried out with instruments
with large effective area at TeV energies, such as ground-based arrays
of Imaging Atmospheric Cherenkov Telescopes (IACTs). The currently
operating IACTs H.E.S.S.\cite{Abdallah:2018qtu}, MAGIC\cite{Aleksic:2013xea} and VERITAS\cite{Archambault:2017wyh}, as well as the Fermi-LAT\cite{Ackermann:2015lka} experiment on satellite and the ground-based water tank array HAWC\cite{Abeysekara:2017jxs}, have done deep
observation campaigns in the Galactic Center (GC) of the Milky Way and
nearby dwarf galaxy satellites of the Milky Way. The next-generation
Cherenkov Telescope Array (CTA) is expected to start data taking
within a decade.

The GC region is arguably the most promising astrophysical environment to detect DM annihilation signals in very high energy (VHE, $E \gtrsim 100 \gev$) gamma rays due to its relative proximity and the expected large accumulation of DM. However, the GC region is  known to be also populated with numerous standard astrophysical emitters in VHE gamma rays. The detection of
sharp spectral features expected from TeV DM annihilations would then be key to provide convincing signatures against the smoother energy spectra exhibited by
astrophysical backgrounds.

The purpose of this work is to improve and update on previous papers\cite{Fowlie:2013oua,Cahill-Rowley:2014twa,Roszkowski:2014iqa,Catalan:2015cna,Aad:2015baa,Beneke:2016jpw,Arbey:2017eos,Athron:2017yua,Abel:2018ekz} that have explored the observational status and prospects of detecting neutralino DM within the  Minimal Supersymmetric Standard Model (MSSM). In our analysis we focus on the upcoming CTA\cite{Acharya:2017ttl} and therefore on the heavy neutralino as DM in the nine-parameter version of the MSSM (p9MSSM) that will be defined below.

Our analysis improves previous works by: (i) deriving the projected CTA sensitivity via a state-of-the-art binned likelihood analysis to be used by the CTA Collaboration, (ii) using up-to-date experimental constraints and numerical tools that include, \textit{e.g.}, 13\tev\ LHC data and \textit{(iii)} taking into account the Sommerfeld enhancement (SE) for all points in the scan, whereas previous works included it only as an estimate or only in some selected sectors of parameter space.

The paper is organized as follows. In Sec.~\ref{sec:DMIDgamma} we
provide an overview of the recent input in VHE gamma-ray results from
the H.E.S.S. experiment in the context of searches for heavy DM. An
update of the CTA sensitivity to DM searches in the GC region using
the latest Monte Carlo simulations of the CTA instrument response
functions is provided. In Sec.~\ref{sec:p9MSSM} we briefly describe
the p9MSSM, scanning methodology and experimental constraints applied
in the analysis. In Sec.~\ref{sec:Results} we compare the results of
our scans with the reach of current and planned indirect and direct
detection experiments. We stress the importance of CTA to provide
coverage of one of the most interesting cases, the $\sim 1\tev$
higgsino region, as emphasized in \cite{Kowalska:2015kaa} (see also\cite{Krall:2017xij,Kowalska:2018toh} for a
recent work and review) that otherwise would remain
unexplored. Finally, we present our conclusions in
Sec.~\ref{sec:Conclusions}.

%%%%%%%%%%%%%%%%%%%%%%%%%%%%%%%%%%%%%%%%%%%%%%%%%%%%%%%%%%%%%%%%%
\section{Indirect detection with VHE gamma rays \label{sec:DMIDgamma}}
%%%%%%%%%%%%%%%%%%%%%%%%%%%%%%%%%%%%%%%%%%%%%%%%%%%%%%%%%%%%%%%%%%

\subsection{Observation of the Galactic Center region with H.E.S.S.}
The GC region is the brightest source of DM annihilation in gamma rays (for a review see \textit{e.g.}\cite{vanEldik:2015qla}).
However, it harbors numerous astrophysical sources that shine in very high energy ($E\gtrsim$ 100 GeV) gamma rays.
Among them are H.E.S.S. J1745-290\cite{Aharonian:2009zk}, a strong emission spatially coincident with the supermassive black hole Sagittarius~A*,
the supernova/pulsar wind nebula G09+01\cite{Aharonian:2005br},
the supernova remnant H.E.S.S. J1745-303\cite{Aharonian:2008gw},
and a diffuse emission extending along the Galactic
plane\cite{Aharonian:2006au,Abramowski:2016mir}.
The rich observational dataset obtained from deep observations
of the GC region by the H.E.S.S. phase-I instrument has been
used to look for continuum and
line signals from DM annihilations. Standard analyses of
H.E.S.S.-I observations of the GC region provided about 250
hours of live time in the inner $1^\circ$ of the GC.

H.E.S.S. searches have been performed with 10 years of data of
the 4-telescope array towards the GC for the
continuum\cite{Abdallah:2016ygi} and mono-energetic
gamma line\cite{Abdallah:2018qtu} channels, respectively, using a 2-dimensional likelihood ratio test statistics to look for any possible excess over the measured background.
In order to avoid modeling the complex standard astrophysical background in the GC region,
a region of $\pm 0.3^\circ$ in Galactic latitude along the Galactic plane has been excluded from the dataset together with
a disk of $0.4^\circ$ radius centered at the position of J1745-303.
No excess in the signal region with respect to background was found and some of the strongest constraints on TeV DM were derived in various annihilation channels\cite{Abdallah:2016ygi,Abdallah:2018qtu}.

\subsection{Dark matter prospects with CTA in the Inner Galactic halo}
The central region of the Milky Way is also a prime target for DM search with the planned CTA\cite{Acharya:2017ttl}.
CTA is envisaged as a two-site observatory to be built at the Paranal site (Chile) in the Southern hemisphere and at La Palma (Spain) in the Northern hemisphere. As the GC region can be observed under favorable and efficient conditions from the Southern hemisphere, the Chilean site of CTA is best suited to explore the GC region. The CTA observation strategy plans a survey of the GC region as a key-science observation program for DM searches\cite{Moulin:2019oyc}.
A deep multi-year observation program is planned in the form of an
extended and homogeneous survey of the inner several degrees of the
GC. The CTA flux sensitivity is expected to improve by up to about one order of magnitude compared to H.E.S.S. and the energy resolution reaches 15\% at about 100 GeV down to better than 5\% in the TeV energy range.
The performance of CTA used in this work is based on instrumental
Monte Carlo simulations performed for the Southern array which
comprises 4 Large-Size Telescopes (LSTs), 25 Mid-Size Telescopes
(MSTs), and 70 Small-Size Telescopes (SSTs). See
Ref.\cite{HASSAN201776} for further details. Following the methodology presented in Ref.\cite{Rinchiuso:2018ajn}, in this work the sensitivity to DM annihilation signals for CTA observations of the GC region is computed using the latest publicly-available instrument response functions (IRFs) of the CTA Southern site at average zenith angle $20^\circ$\cite{CTAIRFs}.

The main background for IACTs  measurements consists of hadronic
(proton and nuclei) cosmic rays (CRs) as well as electron and positron
CRs, with a dominant contribution from the protons. In order to
efficiently separate gamma rays from an overwhelming CR background,
efficient discrimination techniques based of the shower image topology
have been developed\cite{Aharonian:2008zza}. However, due to finite CR
rejection of IACTs, a residual background consisting of
misreconstructed CRs identified as gamma rays is unavoidable. The
expected residual background determination for CTA has been performed
through extensive Monte Carlo simulations\cite{HASSAN201776}. Here, we
use the so-called prod3b version of the instrument response functions
that include the residual background determinations.

The region of interest for the DM signal extraction with CTA extends up to $\pm 5^\circ$ from the GC both in Galactic longitude and latitude.
The overall region is split into squared pixels of side $0.5^\circ$. A homogeneous exposure of 500 hours is assumed over the entire field of view.
The energy-differential residual background rate and acceptance are extracted from Ref.\cite{CTAIRFs}, and the energy threshold is taken at 30\gev. All observations are assumed to be taken at 20$^\circ$ zenith angle.
The IRFs depends on the chosen analysis cuts. All simulations are based on the CTA-South site performance according to an event selection optimized for 50 hours of observation.

The above-mentioned IRFs are provided for {\it on-axis} measurement, {\it i.e.} for emission located near the center of the field of view (FoV). In case of emission distant from the center, the IRFs have been computed as a function of the {\it off-axis} angle and the CTA flux sensitivity has been computed accordingly.
The radius of the FoV region in which the flux sensitivity is within a factor 2 of the one at the center is more than 3 degrees above several hundred\gev\cite{CTAIRFs}. A possible CTA GC survey can make use of a regular grid of pointing positions. Provided that the distance between two
nearby pointings is close enough, an overall spatially homogeneous sensitivity can be obtained. At a few degree distance, the sensitivity reached from a single pointing position degrades significantly but is expected to be compensated by nearby pointings. An optimized and quantitative pointing position strategy for the GC survey with CTA to achieve the best possible sensitivity in the inner several degrees of the GC is much beyond the scope of this work.
In what follows we will assume that an homogeneous flux sensitivity in the overall region of interest with a 500 hour flat exposure can be achieved provided that the overall adequate amount of observation time is granted to the GC survey to fulfill this goal.

\subsection{Statistical method for sensitivity computation}
A dedicated 3-dimensional likelihood ratio test statistics technique has been developed to exploit the spectral and spatial features of the expected DM signal with respect to the background.
The spatial pixels are defined as squared pixels of $0.5^\circ$ between $\pm 5^\circ$ in both Galactic longitude and latitudes. 20 energy bins are taken logarithmically-spaced between energies from 10\gev\ to 100\tev, following Ref.\cite{CTAIRFs}.

The likelihood function for DM searches is defined as a product of the
Poisson probabilities of event counting in the signal and background regions in the $i$-th energy bin, $j$-th Galactic longitude bin, and $k$-th Galactic latitude bin. It reads

\begin{equation}
\mathcal{L}_{ijk}\left(s_{ijk}, b_{ijk}\right)
=
\textrm{Pois}\left(s_{ijk}+b_{ijk}, m_{ijk}\right) \textrm{Pois}\left(\alpha_{jk}b_{ijk}, n_{ijk}\right),
\end{equation}
where the likelihood function follows a Poisson distribution given by
$\textrm{Pois}(\lambda,n) = \lambda^n e^{-\lambda}/n!$ while
$\alpha_{jk}$ corresponds to the ratio of the solid angle size of the
background over the signal regions. The measured count numbers in the
signal and background regions are $m_{ijk}$ and $n_{ijk}$,
respectively.  Following H.E.S.S.' strategy for DM searches in the GC
region, the OFF region (see below) measurements are taken in the same
observational and instrumental conditions as for the signal
measurement, which does not require any further acceptance correction
for the background determination. In this case, $\alpha_{jk}$ is taken
to be~1.  In the context of a counting experiment using {\it ON-OFF}
measurements\cite{Abdallah:2018qtu}, the signal is searched in an ON
region where the measured number of events is $m_{ijk}$.  The expected
number of background events in the signal region, $b_{ijk}$, is
determined from the measurement of number of events in control (OFF)
regions, $n_{ijk}$, with no or little expected searched signal.
$s_{ijk}$ is the expected signal in the signal region.  In order to
compute an expected sensitivity, no excess between the ON and the OFF
regions is assumed, {\it i.e.} $m_{ijk}\equiv n_{ijk}$.where the
likelihood function follows a Poisson distribution given by
$\textrm{Pois}(\lambda,n) = \lambda^n e^{-\lambda}/n!$ while
$\alpha_{jk}$ corresponds to the ratio of the solid angle size of the
background over the signal regions. The measured count numbers in the
signal and background regions are $m_{ijk}$ and $n_{ijk}$,
respectively.  Following H.E.S.S.' strategy for DM searches in the GC
region, the OFF region (see below) measurements are taken in the same
observational and instrumental conditions as for the signal
measurement, which does not require any further acceptance correction
for the background determination. In this case, $\alpha_{jk}$ is taken
to be~1.  In the context of a counting experiment using {\it ON-OFF}
measurements\cite{Abdallah:2018qtu}, the signal is searched in an ON
region where the measured number of events is $m_{ijk}$.  The expected
number of background events in the signal region, $b_{ijk}$, is
determined from the measurement of number of events in control (OFF)
regions, $n_{ijk}$, with no or little expected searched signal.
$s_{ijk}$ is the expected signal in the signal region.  In order to
compute an expected sensitivity, no excess between the ON and the OFF
regions is assumed, {\it i.e.} $m_{ijk}\equiv n_{ijk}$.

The total likelihood function $\mathcal{L}$ is the product of the individual likelihood functions over each $ijk$ bin defined as
${\displaystyle \mathcal{L} = \prod_{ijk} \mathcal{L}_{ijk}}$.
The log-likelihood ratio test statistics (LLRTS) is defined as:
 \begin{equation}
{\rm LLRTS}=-2 \ln\frac{\mathcal{L}\big(s_{ijk}, \hat{\hat{b}}_{ijk})}{\mathcal{L}\big(\hat{s}_{ijk}, \hat{b}_{ijk})}\, ,
\label{TS}
\end{equation}
where the single and double carets indicate unconditional and
conditional maximization, respectively\cite{Cowan:2010js}.
For each mass, a LLRTS is computed and a LLRTS value equal to 2.71 for one degree of freedom corresponds to an one-sided upper limit at 95\%~C.L. on $\sigma v_0$.
The expected sensitivity is computed for a 100\%
branching ratio in each of the channels $W^+W^-$, $ZZ$, $hh$, $Zh$, $c\bar{c}$, $b\bar{b}$,
 $t\bar{t}$, $e^+e^-$, $\mu^+\mu^-$, $\tau^+\tau^-$ and $\gamma\gamma$.

\subsection{Expected signal and background events\label{sec:limits}}
The expected photon flux from pair-annihilation of DM particles of mass $m_{\rm DM}$ in a region of solid angle $\Delta\Omega$ in the sky can be expressed as
\begin{equation}
\label{eq:flux}
\frac{\text{d}\Phi^{\rm DM}_{\gamma}}{\text{d}E}\big(\Delta\Omega,E\big) = \frac{\sigma v_0}{8\,\pi \, m_{\rm DM}^2}\frac{\text{d}N_{\gamma}(E)}{\text{d}E}\, \times \,J\big(\Delta\Omega\big) \, ,
\end{equation}
where $\sigma v_0$ is the total annihilation cross section to all primary channels providing photons in the final sates, and ${\rm d}N_{\gamma}(E)/{\rm d}E$ is the photon spectrum per annihilation.
$J(\Delta\Omega)$ is the so-called J-factor defined as the integral of the square of the DM density $\rho$ along the line of sight $s$ and over $\Delta\Omega$ by
\begin{equation}
\label{eq:jfactor}
J\big(\Delta \Omega\big) \equiv \int_{\Delta \Omega} \text{d}\Omega \int_{0}^{\infty} \text{d}s\, \rho_{\text{DM}} \big(r(s,\theta)\big)^2 \,.
\end{equation}
$s$ is the distance along the line of sight from the observer and is related to the radial distance $r$ in the coordinates centered at GC by $r =  \big(s^2 +r_{\odot}^2-2\,r_{\odot}\,s\, \cos\theta \big)^{1/2}$, where $\theta$ is the angle between the direction of observation and the GC, and $r_{\odot}$ = 8.5 kpc is the distance from the Sun to the GC. We consider a cuspy DM distribution at the GC for which suitable parametrizations are the Einasto\cite{1965TrAlm...5...87E} and Navarro-Frenk-White (NFW)\cite{Navarro:1996gj} profiles defined as
\begin{equation}
\label{eq:profiles}
\rho_{\rm E}(r) = \rho_{\rm s}  \exp \left\{-\frac{2}{\alpha_{\rm s}}\left[\left(\frac{r}{r_{\rm s}}\right)^{\alpha_{\rm s} }-1\right]\right\}\\
\quad {\rm and} \quad \rho_{\rm NFW}(r) = \rho_{\rm s}\left[\frac{r}{r_{\rm s}}\left(1+\frac{r}{r_{\rm s}}\right)^2\right]^{-1},
\end{equation}
where normalization $\rho_s$, scale radius $r_s$, and power index $\alpha_s$ are given in Table \ref{tab:tab1}, following Ref.\cite{Pieri:2009je}.
The local DM density is taken to be $\rho_{\odot} = 0.39\ \rm GeV\ cm^{-3}$\cite{Catena:2009mf}.
Since the DM density in the GC is rather uncertain we also consider a
Cored Einasto profile with a core radius $r_{\rm c}$ such that
$\rho_{\rm CE}(r<r_{\rm c}) = \rho_{\rm E}(r_{\rm c})$ and $\rho_{\rm
  CE}(r\ge r_{\rm c}) = \rho_{\rm E}(r)$. We note that the presence of
possible DM substructures is known to play a subdominant role in DM searches towards the GC and is not therefore considered here (for a discussion see, e.g., Ref.\cite{Gaskins:2016cha} and references therein).

\begin{table}[ht]
\centering
\begin{tabular}{|c|c|c|c|}
\hline
\hline
Profiles & Einasto (E) & NFW & Cored Einasto (CE)\\
\hline
\hline
$\rho_{\rm s}$ (GeVcm$^{-3}$) & 0.079 & 0.307 & 0.079\\
$r_{\rm s}$ (kpc) & 20.0 & 21.0 & 20.0\\
$\alpha_{\rm s}$  & 0.17   &  $-$ & 0.17\\
$r_{\rm c}$ (kpc)  & $-$   & $-$ &   3.0 \\
\hline
\hline
\end{tabular}
\caption{\footnotesize Parameters of the Einasto, NFW and Cored Einasto DM profiles at the GC considered in this work.
\label{tab:tab1}}
\end{table}

The expected DM signal count number in the $ijk$-th bin writes as
\begin{equation}
s_{ijk} = T_{\textrm{obs}} \int_{\Delta E_{i}} dE\, \frac{d\Phi^{\textrm{DM}}_{\gamma}}{d E}\left(\Delta\Omega_{jk},E\right)\,
A_{\rm{eff}}^{\gamma}\left(E\right)\, \mathcal{G}\left(m_{\rm DM}-E\right)
\end{equation}
where $d\Phi^{\rm DM}_{\gamma}/dE$ is defined in \refeq{eq:flux},
$A_{\rm eff}^\gamma(E)$ is the gamma-ray energy-dependent effective area, $T_{\rm obs}$ is the observation time,
and $\mathcal{G}(m_{\rm DM}- E)$ is a Gaussian function centered at the DM mass $m_{\rm DM}$ of width taken as the CTA energy resolution in order to reproduce the effect of the energy resolution
on the theoretical signal spectrum. The DM spectrum $dN_{\gamma}(E)/dE$ is taken from Ref.\cite{Cirelli:2010xx}
for continuum channels.
The monoenergetic gamma-ray line is a Dirac delta function centered at $m_{\rm DM}$.
The signal count rate in the $ijk$-th bin is integrated over the spatial pixel of solid angle size $\Delta\Omega_{jk}$ and energy bin width $\Delta E_{i}$.

The CR background count number is given by
 \begin{equation}
n_{ijk} = T_{\rm{obs}} \int_{\Delta E_{i}}\int_{\rm \Delta\Omega_{jk}}dE\, d\Omega\, \frac{d\Gamma^{\rm CR}}{dE d\Omega}\left(\Omega,E\right),
\end{equation}
where $d\Gamma^{\rm CR}/dE d\Omega$ is the energy-differential residual background rate per steradian.
The background modeling follows the Monte Carlo procedure outlined in the papers, Refs.\cite{APJ2012, HASSAN201776}.

A~detailed modeling of the spectral and spatial extrapolation of the Galactic Diffuse Emission measured by Fermi-LAT in the TeV energy range is beyond the scope of this paper and we neglect it in our computation. A band of $\pm 0.3^\circ$ in Galactic latitudes is excluded from the ROIs as being populated by numerous standard astrophysical sources of VHE gamma rays. A
0.4$^\circ$ radius disk is removed at the position of HESS
J1745-303, one of the brightest TeV gamma-ray sources in the
overall ROI.

%%%%%%%%%%%%%%%%%%%%%%%%%%%%%%%%%%%%%%%%%%%%%%%%%%%%%%%
\begin{figure}[H]
\centering
\subfloat[]{%
\label{fig:linesa}
\includegraphics[width=0.49\textwidth]{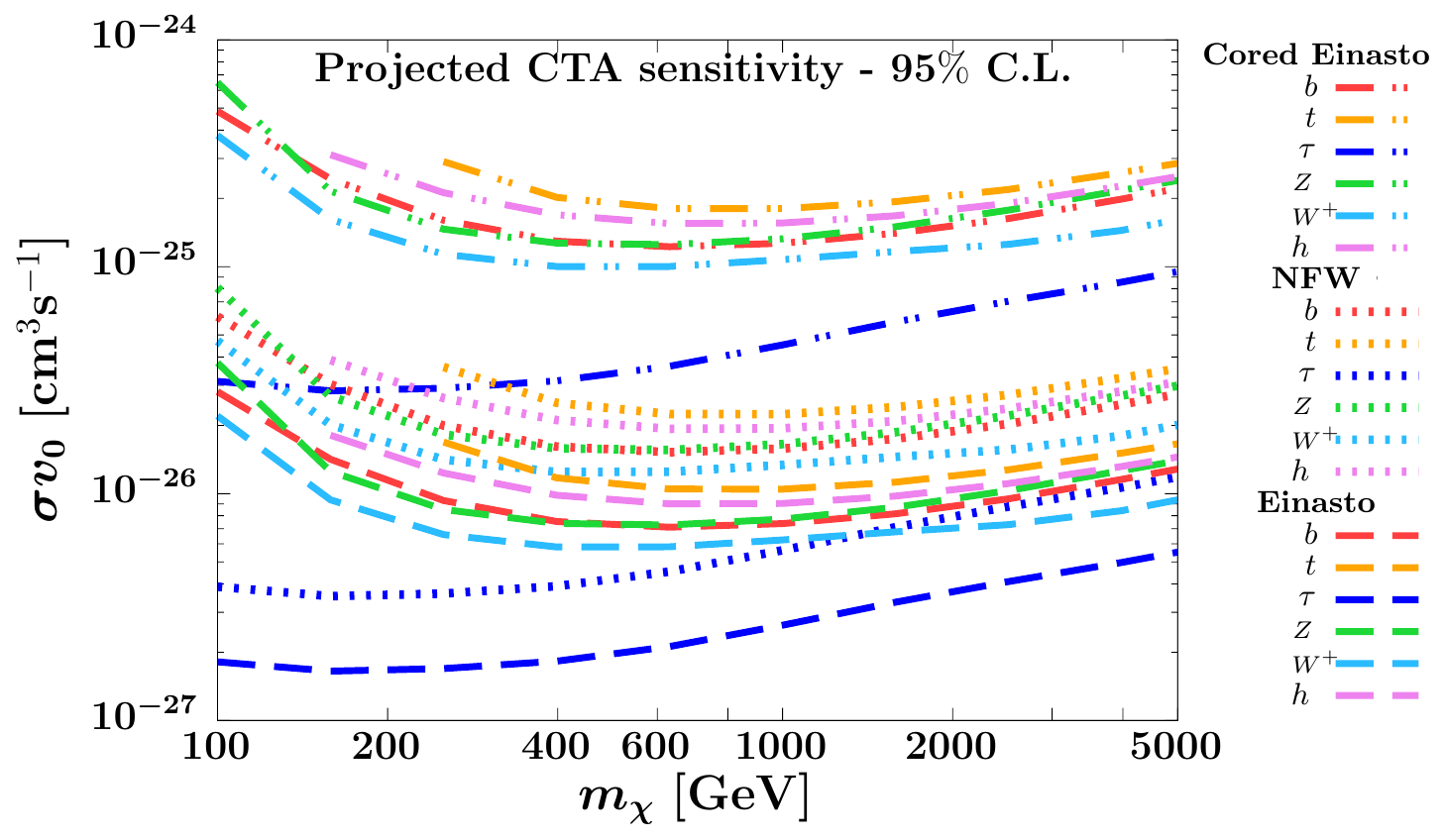}
}%
\hspace{0.02\textwidth}
\subfloat[]{%
\label{fig:linesb}
\includegraphics[width=0.44\textwidth]{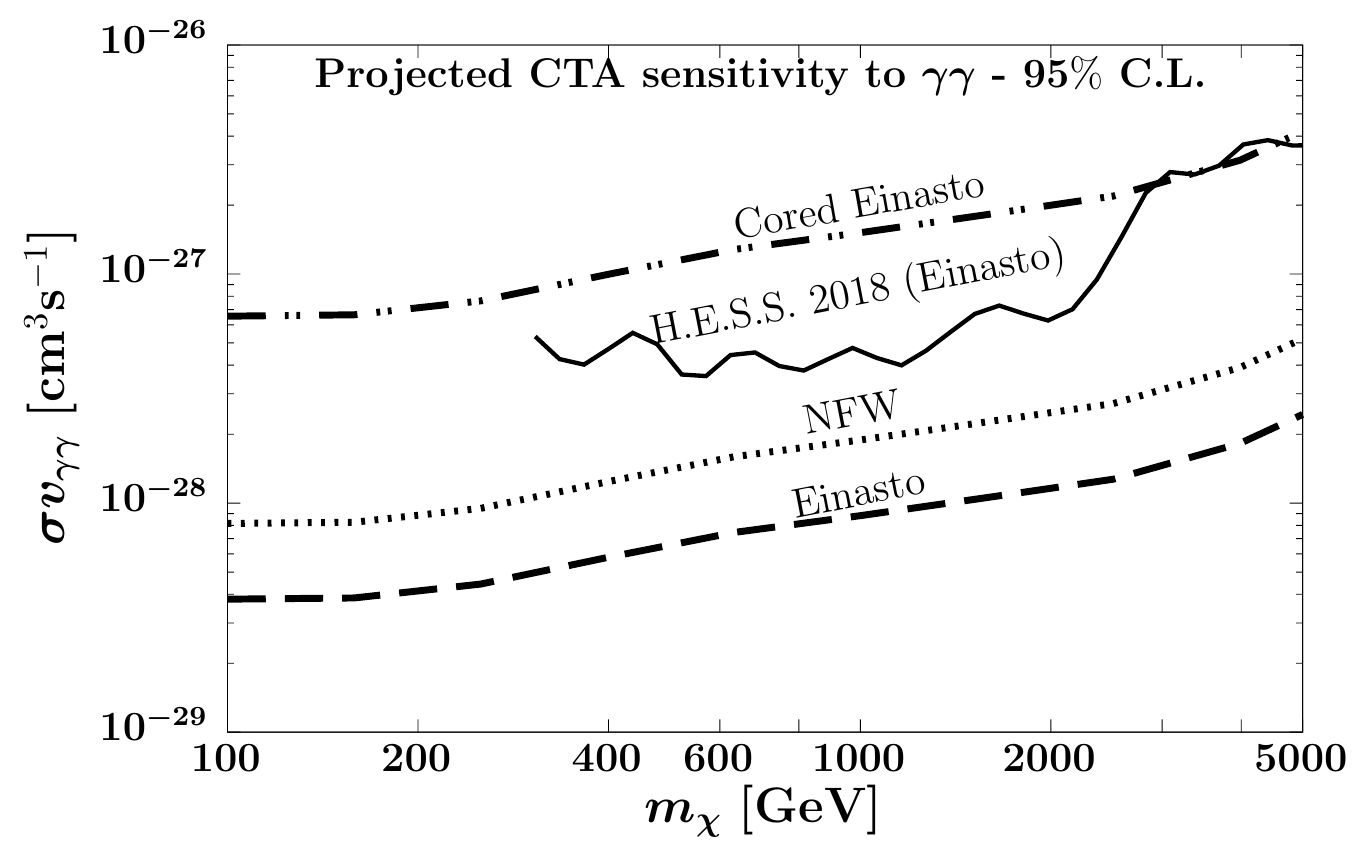}
}%
\caption{\footnotesize 95\%~C.L. CTA projected sensitivities to the velocity-weighted annihilation cross section versus DM mass \mchi,
derived from observations of the inner Galactic halo assuming 500 hour homogeneous exposure
for three separate halo profiles: Cored Einasto (dashed double-dotted), NFW (dotted) and Einasto (dashed lines). (a) The CTA sensitivity is given for the specific final states introduced in the text.
(b) The  CTA sensitivity is derived for a monochromatic $\gamma$ line. The solid line represents the current 95\%~C.L. observed upper limit from H.E.S.S. obtained for the Einasto DM profile\cite{Abdallah:2018qtu}. Line texture is the same as in (a).
}
\end{figure}

%%%%%%%%%%%%%%%%%%%%%%%%%%%%%%%%%%%%%%%%%%%%%%%%%%%%%%%%%%%%%%

We present in \reffig{fig:linesa} \footnote{All sensitivity limits,
  including more annihilation channels, for all three halo profiles
  considered here, can be found in the supplementary material on the arXiv. Limits provided there also extend to 100\tev.}
the projected CTA 95\%~C.L. sensitivity to DM annihilation as a function of DM mass \mchi. For this figure the DM particle is assumed to annihilate into the specific SM final states described in the legend with 100\% branching fraction.
The exclusion lines are computed for each of the three different choices of the DM Galactic halo profile.
The CTA 95\%~C.L. sensitivity to monochromatic $\gamma$-ray lines for the same three choices of halo profile is featured in \reffig{fig:linesb}.
Note that the current monochromatic $\gamma$-ray H.E.S.S. bound, Ref.\cite{Abdallah:2018qtu}, is more constraining than the corresponding Fermi-LAT monochromatic bound, Ref.\cite{Ackermann:2015lka}, in the mass range $m\gsim 300\gev$.

%%%%%%%%%%%%%%%%%%%%%%%%%%%%%%%%%%%%%%%%%%%%%%%%%%%%%%%%%%%%%%%%
\section{The MSSM and details of the scan}\label{sec:p9MSSM}
%%%%%%%%%%%%%%%%%%%%%%%%%%%%%%%%%%%%%%%%%%%%%%%%%%%%%%%%%%%%%%%%
Low energy-scale supersymmetry (SUSY) is the most thoroughly studied scenario for new physics that provides solutions to the problems of the SM - \textit{e.g.} hierarchy problem, lack of DM candidate, unification of gauge interactions. Despite null results for any new physics signal at the LHC or direct and indirect detection experiments searching for DM, SUSY remains an attractive candidate for new physics, especially in light of the discovery at the LHC of a Higgs boson with mass not far above the $Z$ boson mass.
Indeed, the experimental data have so far only excluded models based on optimistic expectations founded on purely theoretical, or aesthetic, arguments, like naturalness.

\subsection{The p9MSSM}
The simplest realization of SUSY that is  also phenomenologically viable is the R-parity conserving MSSM, where the lightest supersymmetric particle (LSP) is stable and may be identified as a thermally-produced DM candidate. We will assume that the LSP is the lightest neutralino.
However, with over 100 free soft-breaking parameters, it is almost
impossible (nor actually even necessary) to study the MSSM in complete generality. Therefore, one has to study more constraining models with a specific high-scale mechanism for SUSY breaking (\textit{e.g.} CMSSM/mSUGRA) or to consider a p(henomenological)MSSM\cite{Djouadi:1998di,Cahill-Rowley:2014twa}. The latter is based on the following assumptions: (i) CP conservation, (ii) Minimal Flavor Violation at the electroweak scale, (iii) degenerate first two generations of sfermion soft-mass parameters and (iv) negligible Yukawa couplings and trilinear couplings ($A$-terms) for the first two generations. In our numerical scan we consider the p9MSSM where in addition to the above, we set the gluino mass, the third-generation down-type right soft squark mass, and the first two generations of soft slepton masses at $20\tev$, which decouples them (see \reftable{tab:p9MSSMv2}). The p9MSSM provides a sufficiently generic parametrization and
coverage of the DM properties of the MSSM with CP and R-parity
conservation. It captures a rich electroweak scale phenomenology with
multiple possibilities regarding its UV-completion, while being
sufficient for our purpose of exploring heavy neutralinos as
DM. Indeed, adding more MSSM parameters to the scan would not alter
our results in any significant way.

\subsection{p9MSSM scanning setup and constraints\label{sec:scan}}
We apply the projected sensitivity reach of CTA as calculated in \refsec{sec:DMIDgamma} to the case of the MSSM parametrized by 9 free input parameters. The parameters that we scan over and their ranges are shown in \reftable{tab:p9MSSMv2}. We employ the $\tt Multinest\_v.2.7$\cite{Feroz:2007kg,Feroz:2008xx} package for the scan, using flat priors. In order to ensure the best coverage of the parameter space of the model, several independent scans with $20,000$ live points each have been performed and the resulting points have been combined when presenting the results.
The supersymmetric spectrum is calculated with \texttt{SPheno v4.0.3}\cite{Porod:2003um,Porod:2011nf}.
We allow the bino mass $M_1$ and the $\mu$ parameter to assume
negative values in order to accommodate blind spots in DM direct
detection\cite{Ellis:2000ds,Cheung:2012qy}, which stem from the
vanishing $h\chi\chi$ coupling for certain combinations of parameters
(see also\cite{Han:2016qtc} for a recent discussion). The remaining
gaugino mass parameter $M_2$ is kept positive, starting from the a
minimal value of 100\gev, allowed by the LEP bounds on charginos. Most
of the third generation sfermion masses are allowed to assume a broad
range of values in between being almost mass degenerate with the
lightest neutralino up to tens of TeV. The former regime allows for
efficient co-annihilations to occur in the early Universe when the DM
relic density is determined, while the latter, in case of squarks, can
more easily lead to a correct value of the Higgs boson mass, $m_h$,
thanks to an increase of the characteristic SUSY scale. As discussed
above, the remaining sfermion mass parameters and the gluino mass
$M_3$ are fixed at 20\tev. They do not play any real role in a further discussion.

\begin{table}[h]
\centering
\begin{tabular}{|c|c|}
\hline
Parameter & Range \\
\hline
\hline
bino mass & $-10 < M_1 < 10$ \\
wino mass & $0.1 < M_2 < 10$ \\
gluino mass & $M_3 = 20$ \\
trilinear couplings & $-30 < A_t=A_b=A_\tau < 30$ \\
pseudoscalar mass & $0.1 < m_A < 10$ \\
$\mu$ parameter & $-10 < \mu < 10$ \\
3rd gen. left soft squark mass & $0.1 < m_{\widetilde{Q}_3} < 30$ \\
3rd gen. right up soft squark mass & $0.1 < m_{\tilde{t}_R} < 30$ \\
3rd gen. right down soft squark mass & $m_{\tilde{b}_R} = 20$ \\
1st/2nd gen. soft squark masses  & $m_{\widetilde{Q}_{1,2}} = m_{\tilde{d}_R,\tilde{s}_R} = 20$ \\
soft slepton masses & $0.1 < m_{\tilde{\tau}_R} = m_{\widetilde{L}_{3}} < 10$ \\
soft slepton masses & $ m_{\tilde{e}_R, \tilde{\mu}_R} = m_{\widetilde{L}_{1,2}} =20$ \\

ratio of Higgs doublet VEVs & $1 < \tan\beta < 62$ \\
\hline
\hline
Nuisance parameter & Central value, error \\
\hline
\hline
Top pole mass \mtop\ (GeV) & (173.34, 0.76)\cite{ATLAS:2014wva}\\
\hline
\end{tabular}
\caption{\footnotesize Ranges of the p9MSSM parameters used in our scans. All masses and trilinear couplings are given in TeV.}
\label{tab:p9MSSMv2}
\end{table}
%%%%%%%%%%%%%%%%%%

The SUSY mass parameters are defined at the scale of the geometrical average of the physical stop masses, $\msusy=(\mstopone\mstoptwo)^{1/2}$. The ratio of the Higgs doublets' vevs, \tanb, and the
top quark pole mass, $m_t$,
which is treated here as a nuisance parameter, are defined at the electroweak symmetry breaking (EWSB) scale. We assume a Gaussian distribution for $m_t$,
whose central value and experimental error are given in\cite{ATLAS:2014wva},
$m_t=(173.34\pm0.76)\gev$.
Our numerical scans are driven by a global likelihood function, which incorporates a standard set of constraints described below.

\paragraph{Dark matter relic density} The constraint with the strongest impact on our numerical result is given by the measurement of the
relic abundance of DM, as given by Planck\cite{Aghanim:2018eyx},
\begin{equation}
\Omega_\chi h^2 = 0.120\pm 0.001.
\label{eq:Oh2Planck}
\end{equation}

To calculate the relic density we employ $\tt micrOMEGAs\ v.5.0.6$\cite{Belanger:2001fz,Belanger:2004yn} supplemented by $\tt DarkSE$\cite{Hryczuk:2011tq}.
We additionally impose a $10\%$ theoretical uncertainty on the calculation to partially take into account the effects of, \textit{e.g.}, loop corrections\cite{Baro:2007em,Baro:2009na}, variations in the renormalization scheme and scale\cite{Harz:2016dql}, and modifications to the QCD equations of state\cite{Hindmarsh:2005ix,Laine:2006cp,Drees:2015exa}.

At the typical mass scale tested by CTA and H.E.S.S. ($\sim 1$ to a few~TeV) the SE plays an important role and strongly affects both the calculation of the present-day neutralino annihilation cross section, $\sigma v_0$, and, to a lesser degree also the determination of the thermal neutralino relic density\cite{Hisano:2006nn,Cirelli:2007xd,Hryczuk:2010zi}. An accurate treatment of the freeze-out process thus requires the incorporation of the SE coming from multiple exchanges of all the gauge bosons and of the SM Higgs and applied to all co-annihilation channels.
At present, the only public code that gives the relic density with the SE included for a generic neutralino and all possible co-annihilation partners in the general MSSM is \texttt{DarkSE} - a package written for \texttt{DarkSUSY v5}\cite{Bringmann:2018lay}.\footnote{\texttt{DarkSE} does not take into account some recent theoretical developments relative to
the most proper way of implementing the SE computation. In particular, the code includes off-diagonal terms in the annihilation matrix\cite{Beneke:2012tg,Beneke:2014gja,Beneke:2014hja} exclusively in the pure wino limit.
In this respect it provides a less accurate determination
than that of a new program that is currently being  developed\cite{Beneke:inprep}, which has been already used in several phenomenological studies\cite{Beneke:2016ync,Beneke:2016jpw}.
However, \texttt{DarkSE} also presents an additional functionality of having the SE implemented for sfermion co-annihilation, which is a necessary ingredient for the scan performed in this work.}

A complete numerical treatment of the SE is very CPU-expensive and thus cannot be handled automatically in a scan. Therefore,
we have adopted a two-step approach: 1) in the scan we use $\tt micrOMEGAs$ and include the SE by rescaling the result using a grid of the enhancements in the $M_2$-$\mu$ plane following the procedure of\cite{Roszkowski:2014iqa};  2) the final points are then post-processed with the accurate SE treatment using full $\tt DarkSE$ code.
Sommerfeld enhancement is also included in the computation of the
present-day $\sigv_0$, as well as for $\sigv_{\gamma\gamma}$ and
$\sigv_{Z\gamma}$.\footnote{It has been checked that the zero-velocity
  limit of these cross sections gives essentially the same result as
  when averaged over the Maxwellian velocity distribution of DM in the GC, with only minimal percent level differences in the close proximity of the SE resonance in the wino region.}
Ideally, one could use an approximate simplified treatments of the SE
for the first step, as in, \textit{e.g.}, \cite{Slatyer:2009vg,ElHedri:2016onc}, but unfortunately these are known only for simple setups - there is no known method for estimating the SE for the relic density with co-annihilations, and there is also no simple functional dependence on either the input nor physical parameters.

Note that in our analysis we do not take into account possible bound-state formation of strongly interacting co-annihilating particles.
This effect was noticed and first discussed for a simple toy model in a recent work\cite{Harz:2018csl} and potentially can apply to the regions of the MSSM parameter space featuring one or more squarks almost degenerate in mass with the neutralino, particularly if the latter lies around the TeV scale.
Implementing bound-state formation in our code would go far beyond the scope of this analysis.
While this effect might modify the value of the predicted relic density for some points, these could only be sporadic cases with very strong co-annihilations with squarks.

Another effect that in principle could have some impact on the discussed limits is the modification of the end point of the energy spectrum of photons produced in the present-day DM annihilation due to soft and collinear gauge boson emission. Such processes, though formally of higher order, are enhanced by large Sudakov logarithms especially at energy scales much larger than the electroweak one. This has been noticed in the context of DM annihilation in\cite{Ciafaloni:2010ti} and explicitly seen in the wino annihilation computation at one-loop\cite{Hryczuk:2011vi} while finally approached with resummation techniques in\cite{Baumgart:2014vma,Bauer:2014ula,Ovanesyan:2014fwa,Beneke:2018ssm,Beneke:2019vhz}.
In\cite{Baumgart:2017nsr,Baumgart:2018yed} it was argued that for the neutralino annihilation the modification due to fully resummed exclusive cross section is most relevant in multi-TeV regime. However, more recently\cite{Beneke:2018ssm,Beneke:2019vhz} showed that for the nearly whole energy regime of interest for CTA, high precision calculations in fact would require resummation of Sudakov logarithms. Nevertheless, due to the fact that it is currently not possible to directly apply the framework of\cite{Baumgart:2017nsr} or\cite{Beneke:2018ssm} to generic p9MSSM neutralinos, and since the expected corrections in the DM mass range of our interest are typically much lower than astrophysical uncertainties, we do not include this effect in our scan.

When performing the numerical scans, we study two commonly discussed cases:
\begin{enumerate}
    \item  the thermal relic density saturates \refeq{eq:Oh2Planck}, in which case we use a Gaussian distribution for the relic density,
    \item the thermal relic density does not exceed the value given in \refeq{eq:Oh2Planck}, in which case we use a half-Gaussian distribution - with relic density imposed only as an upper bound.
\end{enumerate}

In the former case, we assume that no deviations from the standard cosmological history of the Universe took place, as well as that the lightest neutralino is the only DM particle.
In the latter case, the neutralino cannot be a single particle comprising the DM. We then present results of CTA sensitivity to underabundant neutralinos with
local density rescaled by the square of the ratio of the neutralino density to the Planck\cite{Aghanim:2018eyx} value.

On the other hand, the neutralino relic density can also be
significantly affected by deviations from the standard cosmological
history of the Universe, \textit{e.g.}, if neutralino freeze-out
occurs during an extended reheating
period\cite{Giudice:2000ex,Fornengo:2002db,Gelmini:2006pq} (see
also\cite{Roszkowski:2014lga,Drees:2018dsj} for recent studies) or in
presence of additional non-thermal production. In this case, the
neutralino can be a single DM particle even though its standard freeze-out relic density does not saturate the Planck value. In order to accommodate such scenarios, we additionally present results for the aforementioned case 2 but without rescaling $\sigma v_0$.

\paragraph{Dark matter direct detection} The steady progress observed
in recent years in direct detection (DD) searches for DM in
underground liquid noble gas detectors has led to strong upper limits
on the spin-independent cross section of the neutralino scattering off
nucleons.

We include the most recent DD bounds here. For this we employ $\tt SuperIso\ Relic\ v4.0$\cite{Arbey:2018msw} and the $\tt DDCalc\_v.2.0.0$ package\cite{Workgroup:2017lvb}, assuming the Standard Halo Model (SHM) and the following values for the relevant astrophysical parameters:
$\rho_0=0.39~\textrm{GeV}/\textrm{cm}^3$, $v_{\textrm{rot}}=220~\textrm{km}/\textrm{s}$, $v_{\textrm{esc}} = 544\,\textrm{km}/\textrm{s}$.
We note that slight modifications to the SHM that might be suggested by \textit{e.g.} recent data release by the GAIA Collaboration\cite{Brown:2018dum}, see \textit{e.g.}\cite{Necib:2018iwb,Evans:2018bqy} for further discussion, would have minor impact on our results.
The experimental limits that we take into account are the following: PandaX-2\cite{Cui:2017nnn}, PICO-60\cite{Amole:2017dex}, and the most recent results from the XENON1T collaboration\cite{Aprile:2018dbl}.

 Almost universally in the parameter space of the MSSM the bounds on the spin-dependent cross section of the neutralino scattering off the proton or neutron
cannot compete with the corresponding DD bounds on the spin-independent cross section. The current bounds on \sigsdp\ for \mchi\ of our interest come from the searches by the IceCube Collaboration for neutrinos coming from the center of the Milky Way\cite{Aartsen:2017ulx}, the Earth\cite{Aartsen:2016fep}, and the Sun\cite{Aartsen:2016zhm}. Since they are less constraining than the aforementioned DD bounds, and also the indirect detection bounds described below,
we do not consider them here.

In the future, the neutralino scattering cross section on neutrons and protons can also be constrained by their interactions inside neutron stars and white dwarfs\cite{Krall:2017xij,Baryakhtar:2017dbj}. The corresponding limits, however, depend on additional astrophysical assumptions, as well as progress in observations and, therefore, they are not discussed further below.

\paragraph{Collider constraints} The TeV mass-range particle spectrum of the MSSM is very poorly constrained by direct SUSY searches at colliders (see, \textit{e.g.},\cite{Kowalska:2016ent,Arbey:2017eos,Athron:2017yua,Athron:2018vxy} and references therein), including the most recent data from the LHC. In our case, since we focus on the parameter space characterized by colored sparticles lying in the multi-TeV range, only very few points are affected by LHC bounds, with negligible impact on the results shown in~\refsec{sec:Results}.
For completeness, we also take into account LEP and Tevatron limits on SUSY particles\cite{Tanabashi:2018oca}.

\paragraph{Higgs physics} The Higgs mass determination and Higgs-sector LHC measurements
in general can show their effect on the MSSM parameter space under probe in DM searches.
Indirect constraints on the stop mass and mixing from the Higss mass measurement affect the extension of the regions potentially subject to stop co-annihilation; bounds on the mass and couplings of heavy Higgs bosons can
end up influencing somewhat the shape of the funnel regions. In here, the Higgs
sector is constrained with $\tt HiggsBounds-5.2.0beta$\cite{Bechtle:2008jh,Bechtle:2015pma} and $\tt HiggsSignals-2.2.1beta$\cite{Bechtle:2013xfa}, while additional constraints from searches for heavy Higgs decays to $\tau^+\tau^-$ are implemented following\cite{ATLAS:2016fpj,CMS:2016rjp}.

\paragraph{Flavor physics} We calculate a few flavor observables with $\tt Superiso\ Relic\ v4.0$\cite{Arbey:2018msw}.
The parameter space of the MSSM is potentially sensitive, in particular, to the bounds from rare decays in $b\to s ll$ processes and
radiative decays like $b\to s \gamma$, which can constrain scan points characterized by large \tanb\ values,
and/or relatively light non-SM Higgs bosons, squarks, and charginos/neutralinos.
We use the following experimental determinations:
\bea
\textrm{BR}\left( B \rightarrow X _ { s } \gamma \right) &=& ( 3.27 \pm 0.14 ) \times 10 ^ { - 4 },\label{bsgamma}\\
\textrm{BR}\left( B _ { s } ^ { 0 } \rightarrow \mu ^ { + } \mu ^ { - } \right)& =& \left( 3.0 \pm 0.6 _ { - 0.2 } ^ { + 0.3 } \right) \times 10 ^ { - 9 }\label{bsmumu}
\eea
where, following, \textit{e.g.}, Ref.\cite{Workgroup:2017myk}, in \refeq{bsgamma} we give the
calculated average\cite{Misiak:2017bgg} of the determinations in
Refs.\cite{Aubert:2007my,Lees:2012wg,Lees:2012ym,Saito:2014das,Belle:2016ufb}, and in
\refeq{bsmumu} we report the most recent LHCb measurement, based on 8\tev\ collision
data\cite{Aaij:2017vad}. We thus implicitly assume that \refeq{bsmumu} has superseded
an older statistical combination of CMS and LHCb measurements with 7 and 8\tev\
data\cite{CMS:2014xfa}. Note that very recently the ATLAS Collaboration has presented a measurement of
$\textrm{BR}\left( B _ { s } ^ { 0 } \rightarrow \mu ^ { + } \mu ^ { - } \right)$, from a combination of data taken during their 8\tev\ and 13\tev\ runs,
which agrees with \refeq{bsmumu}: $\textrm{BR}\left( B _ { s } ^ { 0 } \rightarrow \mu ^ { + } \mu ^ { - } \right)=(2.8_{ - 0.7 }^{ + 0.8 })\times 10^{ -9}$\cite{Aaboud:2018mst}.

We impose the bounds of Eqs.~(\ref{bsgamma}) and (\ref{bsmumu})
at the 95\%~C.L., \textit{a posteriori} on the points belonging to the $2\,\sigma$ region of the profile likelihood. This
reduces the number of viable points in the scan by approximately 2\%. Other potentially relevant flavor observables like $\textrm{BR}(B^{\pm}\to\tau\nu_{\tau})$ or the $B_s$ mass mixing measurement $\Delta M_{B_s}$
are not constraining at the mass scale relevant for this paper.

Note that we do not include constraints from observables that are currently
showing a 2--3~$\sigma$ discrepancy with the SM,
like the differential branching ratios and angular observables in
$B^0\to K^{\ast 0}\mu^+\mu^-$\cite{Aaij:2014pli,Aaij:2015oid}, or the branching ratio measurements
that have recently provided tantalizing hints of lepton flavor
nonuniversality\cite{Lees:2013uzd,Aaij:2015yra,Hirose:2016wfn,Aaij:2017vbb,Aaij:2019wad}.
It is known that these
anomalies cannot be explained consistently in the MSSM (see, \textit{e.g.},
Ref.\cite{Altmannshofer:2014rta}) and that, if confirmed to higher statistical significance with further release of data,
will require new physics beyond the particle content of the MSSM.
For analogous reasons, we do not apply to the parameter space the constraint from the measurement\cite{Bennett:2006fi} of the
muon anomalous magnetic moment, which shows a $3.5\,\sigma$ discrepancy with the SM expectation, $\deltagmtwomu=(27.4\pm 7.6)\times 10^{-10}$\cite{Davier:2016iru}.
It is well known that this value cannot be accommodated in the regions of the MSSM parameter space that feature
a TeV-scale LSP, see, \textit{e.g.}, Ref.\cite{Kowalska:2015zja}. In this case too, if the anomaly were to be confirmed by upcoming Fermilab
data\cite{Grange:2015fou}, it will require a BSM explanation lying outside
of the MSSM parameter space relevant for the current analysis.
One should keep in mind, however, that is possible
to extend the MSSM minimally by a U(1) gauge group, so that
the all of the above-mentioned flavor anomalies become consistent with TeV-scale neutralinos with the
exact same DM properties as in the vanilla MSSM\cite{Darme:2018hqg}.

\paragraph{Dark matter indirect detection} The indirect detection constraints on neutralino DM, that are the main subject of this study, are not included in the likelihood function when performing initial numerical scans of the parameter space of the MSSM. Instead, we carefully study them by postprocessing the results obtained in these scans. This leads to a better understanding of their impact on the allowed parameter space.

The most constraining data for the TeV-scale mass range are currently provided by H.E.S.S. A more detailed description of DM ID limits from H.E.S.S. and future projections has been described in details in \cref{sec:DMIDgamma}.

When presenting the results below, we also take into account Fermi-LAT limits on DM-induced $\gamma$-rays that correspond to 6~years of data and observation of $28$ dSphs\cite{Fermi-LAT:2016uux}. These data are in principle most constraining in the MSSM for neutralinos of mixed gauge composition with a mass of a few hundred~GeV, which are, however, already strongly bounded by the null DD results. They might also provide a complementary probe on the low-energy tail of spectra from the annihilation of winos including SE. We illustrate this below for a fixed annihilation final state into a $b\bar{b}$ pair. We have also verified numerically, following $\tt Superiso\ Relic\ v4.0$\cite{Arbey:2018msw}, that taking into account a complete list of annihilation final states leads to similar results.

We note that \sigv\ can also be constrained by requiring that the CMB spectrum is not affected too much by the pre- and
post-recombination energy injection from DM annihilations\cite{Padmanabhan:2005es,Galli:2009zc,Slatyer:2009yq}.
However, for the heavy DM of our interest, this effect typically leads to less stringent bounds than null searches for
DM annihilation signal in the GC by H.E.S.S. and in dSphs by Fermi-LAT (see\cite{Beneke:2016jpw} for recent discussion).

A recent determination\cite{Cuoco:2017iax} of the bounds on the neutralino annihilation cross section from AMS-02 antiproton cosmic-ray (CR) data\cite{Aguilar:2016kjl} has proven to be competitive with H.E.S.S. diffuse $\gamma$-ray searches in the $\sim{\rm TeV}$ mass range.
We have verified that this is in general true also in the context of our scans using $\tt SuperIso\ Relic\ v4.0$\cite{Arbey:2018msw} that employs a semi-analytic approach to solving the propagation equations following\cite{Boudaud:2014qra}. However, the limits that one derives from the AMS-02 data depend on the assumed CR propagation model and suffers from large astrophysical uncertainties (see, \textit{e.g.}, Refs.\cite{Evoli:2011id,Cirelli:2013hv}). For this reason, we do not discuss them in details in the following section,
which focuses on DM-induced $\gamma$-ray signal.

%%%%%%%%%%%%%%%%%%%%%%%%%%%%%%%%%%%%%%%%%%%%%%%%%%%%%%%%%%%%%%%%%%%%%%
\section{Results}\label{sec:Results}
%%%%%%%%%%%%%%%%%%%%%%%%%%%%%%%%%%%%%%%%%%%%%%%%%%%%%%%%%%%%%%%%%%%%%%
\subsection{CTA sensitivity to the p9MSSM parameter space}

For each point in the scan, we compute the H.E.S.S. limit for the present-day annihilation cross section, $\sigma v_0$, and the corresponding projected sensitivity of CTA. We use the 95\%~C.L. bounds and projections
for annihilation to pure channels  (see \reffig{fig:linesa}). In case of annihilation final states for which H.E.S.S. limits have not been reported by the collaboration, we employ the most relevant existing bounds. In particular, for $hh$ final state we use $ZZ$ limit, for final states with $c$ and $s$ quarks -- $b\bar{b}$ limit, for the lightest quarks -- $\tau^+\tau^-$ limit and for $e^+e^-$ we employ $\mu^+\mu^-$ limit. Instead, for CTA we derive bounds for a more complete set of annihilation final states, as discussed in section~\ref{sec:limits}.

In order to verify whether a particular point in the p9MSSM parameter space is within current bounds and future sensitivities, we combine limits obtained for pure annihilation final states by taking their average weighted by the branching-ratios corresponding to those channels.
In section~\ref{sec:benchmarks} we show that this procedure is sufficient for our purpose, by comparing our results for several benchmark scenarios to a more detailed treatment in which photon spectra are carefully combined prior to obtaining the CTA limit.

For channels with non-SM particles in the final state, \textit{e.g.}, the neutral MSSM Higgs particles, $A_0$ and $H$, we employ the bounds computed for the SM Higgs; for the charged MSSM Higgs particle, $H^{\pm}$, we use the bounds derived for $W^{\pm}$.
While the non-SM annihilation final states typically do not play a dominant role in our analysis, they might become important for selected points in the parameter space. For these points, we have verified our results against a more detailed procedure in which decays of the non-SM particles were taken into account employing $\tt HDECAY$\cite{Djouadi:1997yw, Djouadi:2018xqq} before generating the combined photon spectrum using Ref.\cite{Cirelli:2010xx}.

In the plots below we only show points that belong to the 95\%~C.L. region of the global profile-likelihood, i.e. we select $\Delta \chi^2 \leq 5.99$ from the best-fit point, where $\Delta\chi^{2}=-2 \ln \left(\mathcal{L} / \mathcal{L}_{\max }\right)$.

%%%%%%%%%%%%%%%%%%%%%%%%%%%%%%%%%%%%%%%%%%%%%%%%%%%%%%%%%%%%%%%%%%%%%%%%%%%%%%%%%%%%
\subsection{Discussion of results\label{sec:results}}

\begin{figure}[t]
\centering
\includegraphics[width=1.0\textwidth]{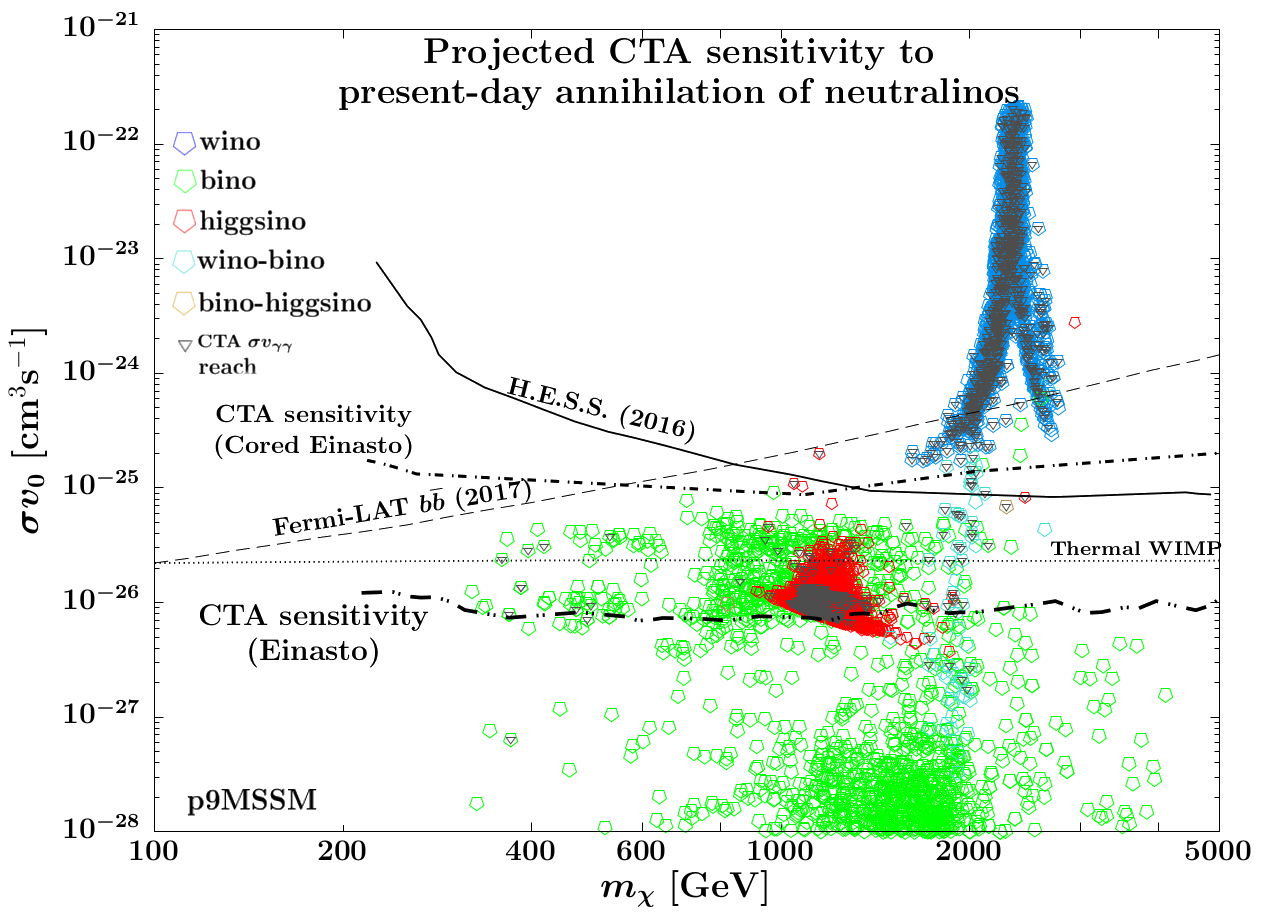}
\caption{\footnotesize Distribution of p9MSSM points with $\Delta
  \chi^2\leq 5.99$ in the (\mchi, $\sigma v_0$) plane. The color
  coding reflects the composition of the lightest neutralino, as
  discussed in the text and according to the legend. The current upper
95\%~C.L. limits from H.E.S.S.\cite{Abdallah:2016ygi} applied to the p9MSSM are indicated with a black solid line.
The projected CTA sensitivity applied to the p9MSSM is shown as a thick (Einasto), or thin (Cored Einasto) dashed double-dotted line.
All points above the line will be probed at the
$\sim$95\%~C.L.
The Fermi-LAT\cite{Fermi-LAT:2016uux} $b\bar b$ mode from dwarf spheroidal galaxies
is shown as a dashed line.
To highlight the complementarity between the continuous and
monochromatic photon search, we denote the points whose $\sigma
v_{\gamma\gamma}$ is within reach (assuming Einasto halo profile) at
CTA by dark gray triangles.
\label{fig:pMSSM1}}
\end{figure}

We present in \reffig{fig:pMSSM1} the scan points
in the plane (\mchi, $\sigma v_0$) of the present-day annihilation cross section of the neutralino versus its mass.
The color code used in \reffig{fig:pMSSM1} refers to the gauge composition of the neutralino LSP, which, by construction, in the MSSM is never a 100\% pure eigenstate.

How ``pure'' a certain mass eigentate is depends
on the elements of the unitary matrix, $\boldsymbol{Z}$, diagonalizing the neutralino mass matrix after EWSB.
In green we show the points with the LSP containing at least 90\% of
the pure bino gauge eigenstate (\textit{i.e.}, that is,
in the basis of gauge eigenstes \{bino, wino, down-type higgsino, up-type higgsino\}, we require $|Z_{11}|^2>0.9$ for these points).
In blue, the points for which it is for at least 90\% a wino ($|Z_{12}|^2>0.9$).
In cyan we show a mixture of these two gaugino states, with the additional constraint that the higgsino composition remain below 10\%, $|Z_{13}|^2+|Z_{14}|^2<0.1$.
In red we show neutralinos that are dominated for at least 90\% by their higgsino fraction ($|Z_{13}|^2+|Z_{14}|^2>0.9$).
We finally point out that only very few points characterized by a
mixture of a gaugino and a large higgsino component appear, marked in gold,
in the plot, as they are in strong tension with the latest bounds from direct detection searches.

We also note that due to our general focus on TeV-scale neutralino DM, which is of  most relevance for H.E.S.S and CTA, our scanning procedure is not tuned to thoroughly explore the parameter space of the p9MSSM corresponding to light neutralinos with masses around
the EWSB scale. For this reason, we do not show in our plots points corresponding to the region where $m_\chi\approx m_h/2$ where
the correct neutralino DM relic density can be obtained thanks to efficient resonance annihilations via the Higgs boson exchange.
We note, however, that the expected annihilation cross section for such light neutralino DM lies well below the reach of CTA.
The same is also true for another instance of supersymmetry at the electroweak scale that has recently been discussed in the context of a collection of mild excesses present in the LHC data\cite{Athron:2018vxy}.

We show in \reffig{fig:pMSSM1} with a solid black line the current
95\%~C.L. upper bound on $\sigma v_0$ from 254~hours of observation of
the GC at H.E.S.S., under the Einasto profile assumption, applied to
the points of the p9MSSM. Importantly, when deriving these results, as
well as CTA sensitivity lines discussed below, we take into account
all the points obtained in the scan of the parameter space including
the ones that violate some of other constraints and, therefore, are
not shown in the plot. In particular, the presence of such otherwise
excluded points allows us to determine the position of the H.E.S.S. limit in the region with a light neutralino and large $\sigma v_0$ which is virtually excluded by current bounds.

The latest observations exclude points whose neutralino is strongly dominated by the wino component (in blue, and some in cyan),
for which the annihilation cross section in the present day has a large SE\cite{Cohen:2013ama,Hryczuk:2014hpa}.
The plot updates Figure~5(a) of Ref.\cite{Roszkowski:2014iqa} and is in agreement with {\it e.g.} Ref.\cite{Catalan:2015cna}.
Compositions of the neutralino very close to a pure wino state are in very strong tension with H.E.S.S. with continuum observations
as well as monochromatic line searches.

The H.E.S.S. bound, on the other hand, does not bite into the $\sim1\tev$ (nearly pure) higgsino region of the parameter space,
corresponding to the red points in Fig.~\ref{fig:pMSSM1}, for which the SE is less pronounced.
Upcoming increased statistics can tighten the bound but, realistically, batches of new data are at this point not expected to bring qualitative improvements to the current picture.
It is CTA, with an effective area that is by about a factor 10 larger
than that of H.E.S.S.' at 1~TeV, that will be probing more deeply into the higgsino region of the parameter space.

We show with a dash-dotted black line our projection of the sensitivity of CTA in the p9MSSM in searches for DM-induced diffuse photon flux,
with 500 hours of observation of the GC and under the Einasto profile assumption.
In addition, we overlap gray triangles to the
points that are within the sensitivity of the CTA $\gamma$-ray line search.
As was described in \refsec{sec:DMIDgamma}, we factor in a detailed treatment of the statistical uncertainties, and the likelihood function is calculated with an improved design of the ROIs of the Galactic Plane with respect to previous analyses\cite{Roszkowski:2014iqa,Lefranc:2015pza}.
The higgsino region of the parameter space is likely to be tested in its near entirety by CTA, and the same is true for points with bino-dominated neutralinos
with annihilation cross section around the thermal freeze-out value (light dotted line).

We note that the actual CTA limit on the p9MSSM parameter space cannot be perfectly represented by a single line due to number of possible neutralino DM annihilation final states that need to be taken into account. However, we have verified that, for $m_\chi\gtrsim 1$~TeV the approximate limit that we present reproduces very well the true CTA sensitivity. For lower masses, the line shown in \reffig{fig:pMSSM1} corresponds to a conservative approach, \textit{i.e.}, all the points lying above the line are within the CTA reach. We follow a similar strategy to obtain the approximate H.E.S.S. limit shown as a solid line in \reffig{fig:pMSSM1}.

\begin{figure}[ht]
\centering
\includegraphics[width=0.9\textwidth]{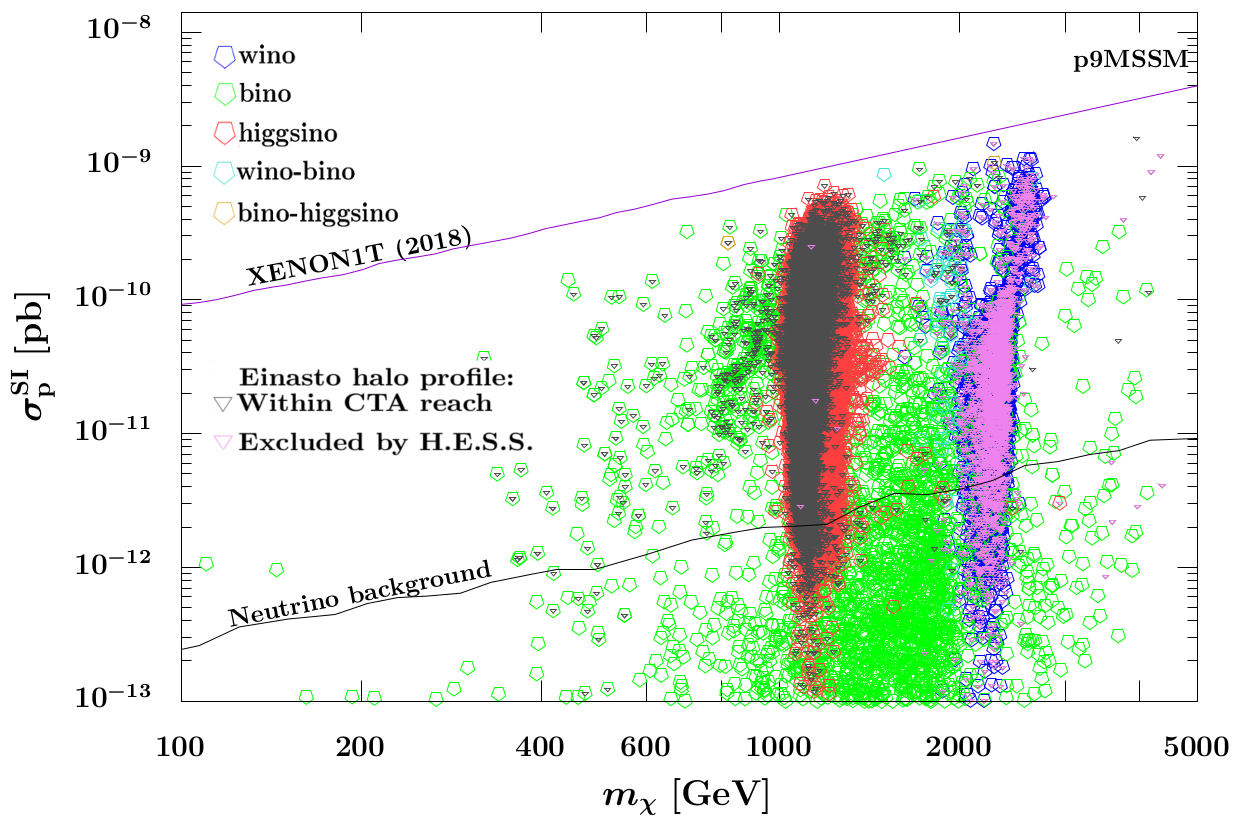}
\caption{\footnotesize Distribution of p9MSSM points with $\Delta \chi^2\leq 5.99$ in the (\mchi, \sigsip) plane with color coding as in \reffig{fig:pMSSM1}. Points excluded by H.E.S.S. (Einasto, both continuous and monochromatic photon search) are denoted by violet triangles, while those
within the sensitivity of CTA (Einasto, both continuous and monochromatic photon search) are denoted by black triangles.
The most recent limit from the XENON1T Collaboration\cite{Aprile:2018dbl}, which is included in the likelihood function, is denoted by a purple solid line, while the onset of the irreducible neutrino background is denoted by a black solid line.
\label{fig:pMSSM2}}
\end{figure}

The points that will remain untested, almost all characterized by a nearly pure 
bino-like composition of the LSP, are those for which the neutralino annihilation
cross section is too small to yield the correct relic density, and thus either feature spectra with sparticles that co-annihilate in the early Universe with
the LSP (near mass degeneracy between the bino-like neutralino and one
or more sfermions), or spectra that include one or more Higgs bosons of mass
within a few hundred GeV of $2\mchi$, which provide a means for \textit{funnels}, or resonant $s$-channel annihilation of the LSP in the early
Universe due to the thermal broadening of the energy distribution.
As is well known, the specifics of these spectra are very model-dependent. Moreover, their realization in explicit high-scale completions can encounter model building challenges and/or require some fine tuning of the initial parameters.

This is unlike in the case of (nearly pure) higgsinos and winos, which do fall inside the sensitivity of large IACTs,
and for which the correct value of the relic density emerges naturally
once the mass of the LSP is around either 1\tev, or $\sim 2.5-3\tev$, respectively, quite independently  of the model details of the rest of the sparticle spectrum.
Note, however, that for higgsino points with masses larger than about 1.3\tev, shown outside of the CTA sensitivity in \reffig{fig:pMSSM1}, one also relies on additional mechanisms like coannihilations with squarks in the early Universe to preserve the correct relic density. These points tend to feature lower present-day annihilation cross section than lighter higgsinos, and they are consequently more difficult to probe.

The projected sensitivity of CTA shown in \reffig{fig:pMSSM1} is obtained in the two limiting cases of Einasto and Cored Einasto DM halo profiles.
A sensitivity line corresponding to the NFW  profile can be easily obtained by multiplying the projected line for Einasto case by the factor of about 2.5 obtained from \reffig{fig:linesa}.

In \reffig{fig:pMSSM1}, we also show with the dot-dashed line the projected CTA sensitivity reach obtained for the Cored Einasto profile defined in \refsec{sec:limits}. As can be seen, in this case CTA can still play an important role by probing the entire wino-like neutralino DM scenario which would otherwise remained not fully tested by the H.E.S.S. observations.

In \reffig{fig:pMSSM2} we show the p9MSSM points in the (\mchi, \sigsip) plane. The most recent XENON1T 90\%~C.L. upper limit\cite{Aprile:2018dbl} is
shown by a solid violet line. The XENON1T results are included in the global likelihood function, and that explains the absence
above the line of any point belonging to the $\sim 2\sigma$ region of the profile likelihood.
The onset of the irreducible neutrino background is denoted by a solid black line.
The color code is the same as in \reffig{fig:pMSSM1} and we additionally overlap violet triangles to points excluded by the H.E.S.S.
bound on $\sigma v_0$. Black triangles are overlapped to points within our projection of the sensitivity of CTA in the Einasto profile.

\begin{figure}[ht]
\centering
\includegraphics[width=0.9\textwidth]{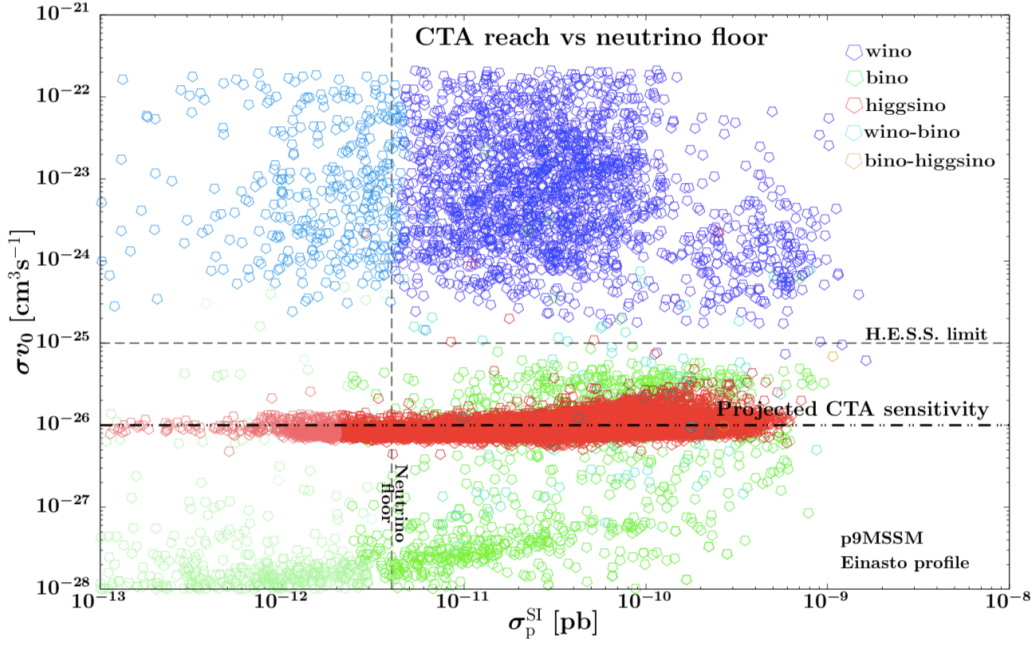}
\caption{\footnotesize The points of the p9MSSM in the (\sigsip, $\sigma v_0$) plane. Color coding is the same as in \reffig{fig:pMSSM1}. The upper limit on \sigsip\ is determined by the current sensitivity of XENON1T, included in the global likelihood function.
The shaded region covers the points lying below the irreducible neutrino floor. To guide the eye we add a vertical dashed line with reference value of the neutrino background limit taken at $\mchi=2\tev$. The dashed horizontal line denotes the H.E.S.S. 95\%~C.L. upper limit, taken at $\mchi\approx 2.5\tev$, while the dashed double-dotted horizontal line denotes the approximate CTA reach, taken at $\mchi\approx 1\tev$.
\label{fig:pMSSM3}}
\end{figure}

The necessity of using both direct and indirect detection strategies to cover the most substantial portions of the parameter space of
the MSSM with high-mass DM has been pointed out in the literature since early after the discovery of the Higgs boson at the LHC\cite{Roszkowski:2014iqa}.
We show the power of complementarity of direct and indirect detection in \reffig{fig:pMSSM3},
where we project the points of the p9MSSM to the (\sigsip, $\sigma v_0$) plane. The color code is the same as in \reffig{fig:pMSSM1} and \reffig{fig:pMSSM2}.

The future reach of direct underground searches with noble liquids is bound to bite into the parameter space from right to left, until it reaches
the  irreducible neutrino background, shown here as a shaded region
(recall that the value of \sigsip\ characteristic of the neutrino ``floor''
for direct DM searches depends on the DM mass, hence the boundary of the shaded area is jagged in \reffig{fig:pMSSM3}).
To guide the eye, we add a vertical dashed gray line denoting the neutrino background limit $\sigsip\approx 4\cdot10^{-12} \ \text{pb}$ taken at $\mchi\approx 2\tev$.
Conversely, the sensitivity of IACTs gradually improves from the top down, providing a complementary means of testing the parameter space. The H.E.S.S. bound
is denoted in the figure by a dashed black  horizontal line while the projected sensitivity of CTA is denoted by a dashed double-dotted horizontal line.

%%%%%%%%%%%%%%%%%%%%%%%%%%%%%%%%%%%%%%%%%%%%%%%%%%%%%%%%%%%%%%%%%%%%%%
 \subsection{Underabundant neutralinos}

\begin{figure}[htb]
\centering
\subfloat[]{%
\label{fig:3a}
\includegraphics[width=0.45\textwidth]{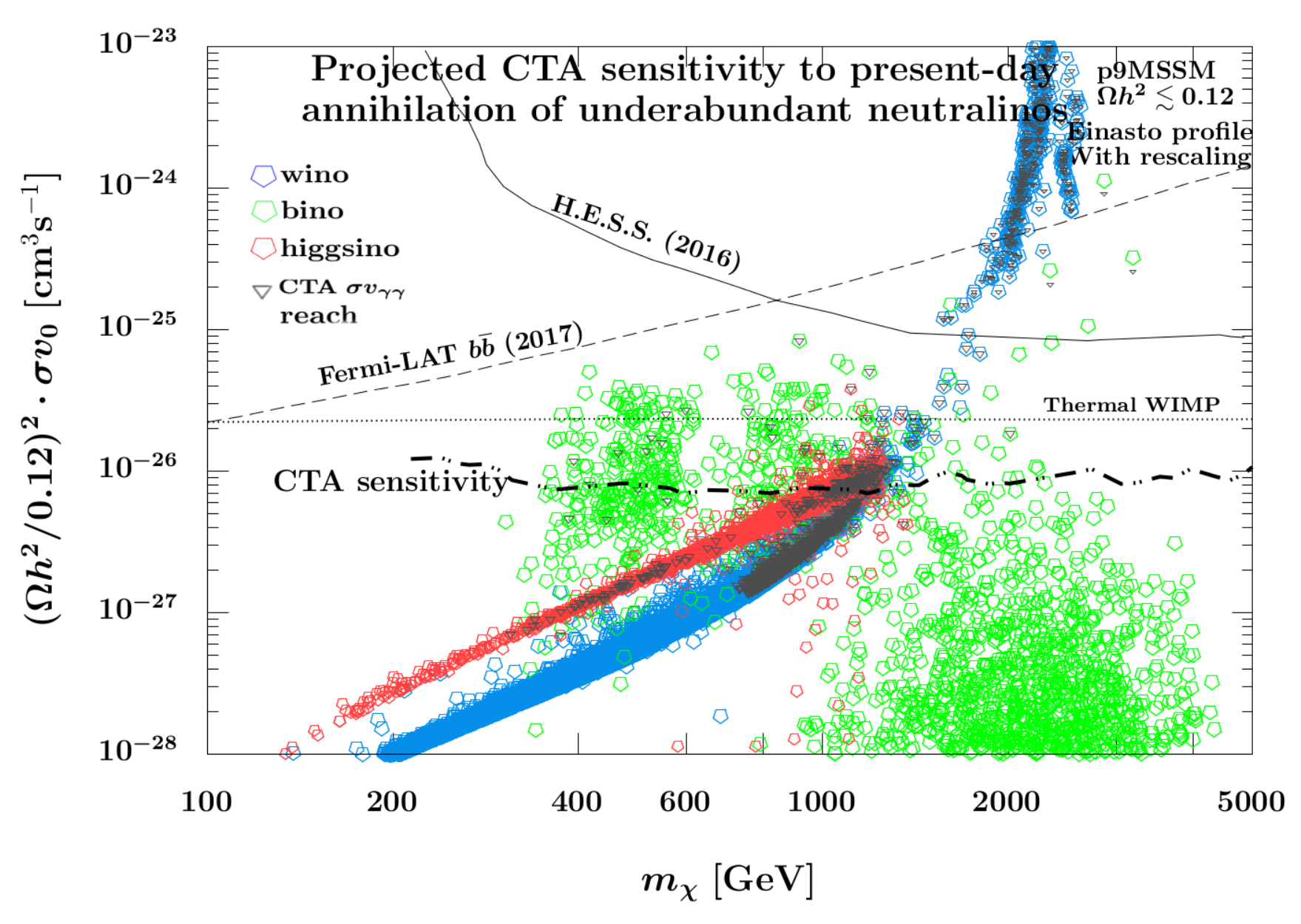}
}%
\hspace{0.02\textwidth}
\subfloat[]{%
\label{fig:3b}
\includegraphics[width=0.45\textwidth]{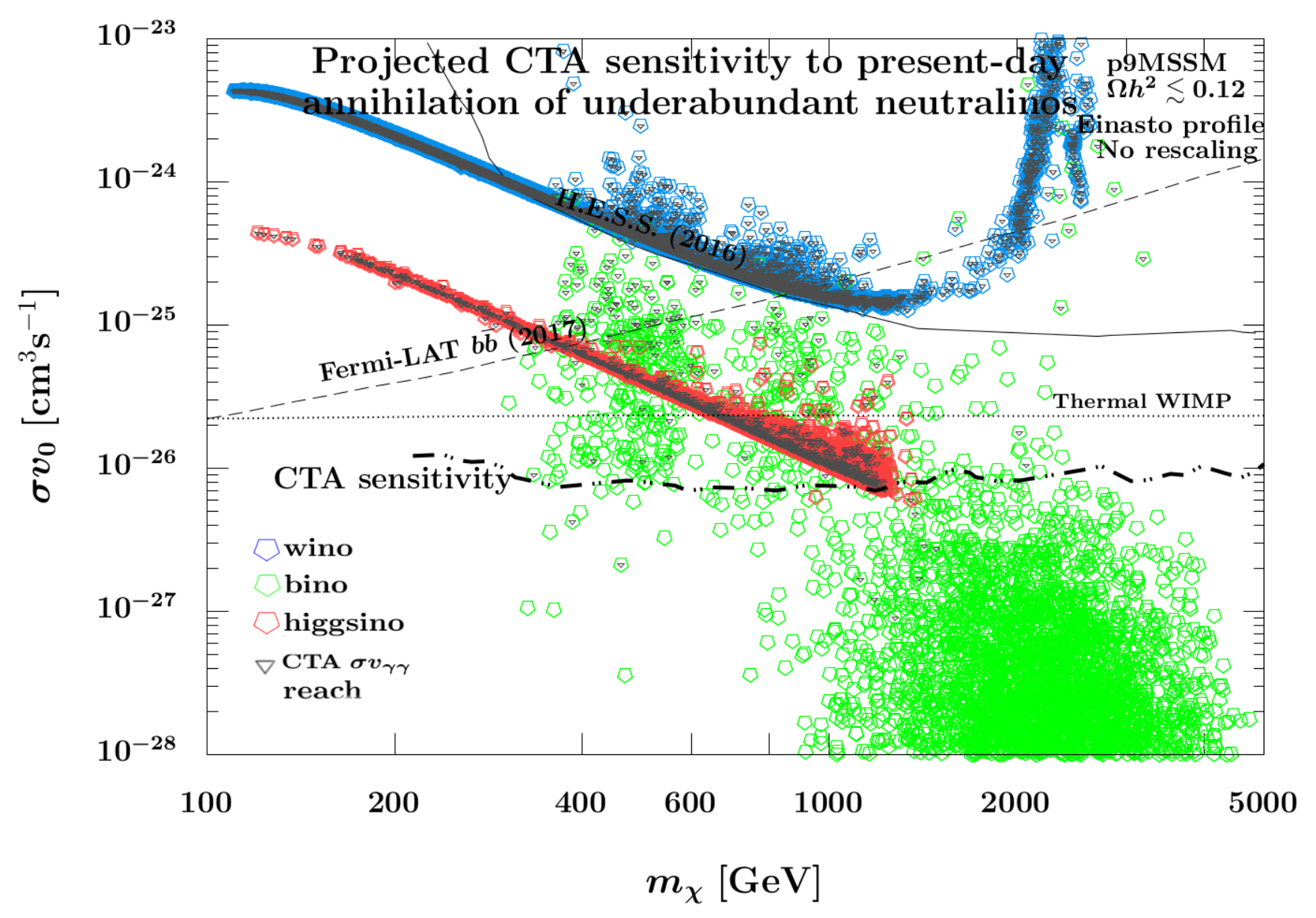}
}%
\caption{\footnotesize (a) Distribution of points with $\Delta
  \chi^2\leq 5.99$ in (\mchi, $(\Omega_{\chi} h^2/0.12)^2\cdot\sigma
  v_0)$ plane for underabundant neutralinos. Color and line texture
  coding is the same as in \reffig{fig:pMSSM1}. (b) Same as in (a) but
  without rescaling.}
\end{figure}

As discussed in Sec.~\ref{sec:scan}, the neutralino can be a good DM candidate even when its thermally produced relic abundance is different from the total DM relic density in the Universe. It can then either be one of several DM components, or might even remain the only DM particle but in non-standard cosmological scenarios. In this subsection, we present the results of two scans corresponding to the cases in which the relic density constraint is imposed as an upper bound only, by means of a half Gaussian distribution. The corresponding results can be seen in Figs.~(\ref{fig:3a}) and (\ref{fig:3b}) where only the points that belong to the 95\%~C.L. region of the global profile-likelihood are shown in the (\mchi, $\sigma v_0$) plane.

In \reffig{fig:3a} we rescale $\sigma v_0$ by $(\Omega_{\chi} h^2/0.12)^2$ which corresponds to the case when neutralino DM can provide only a partial contribution to the total $\Omega_{\textrm{DM}}h^2$. Similarly, we rescale the DM DD cross section \sigsip\ by $\Omega_\chi h^2/0.12$ when imposing the corresponding constraints. As can be seen in the plot, underabundant higgsino-like and wino-like neutralinos with masses of order few hundred GeV are typically beyond the reach of CTA. There are, however, some higgsino-like points that can be probed by the CTA monochromatic photon ($\sigv_{\gamma\gamma}+\frac 12 \sigv_{\gamma Z}$) search even though these points lie below the projected CTA sensitivity in searches for DM-induced diffuse photon spectrum (dash-dotted line in the plot). These points are denoted by gray triangles. The crucial impact of the monochromatic line search is even more pronounced for heavier neutralinos with masses $m_\chi\approx 1\tev$. In particular, it is worth stressing that, \textit{e.g.}, an underabundant wino-like neutralino DM with $m_\chi\approx 1\tev$ can be discovered by CTA in monochromatic-line searches with no corresponding signal in the diffuse spectrum searches. For even heavier, but still underabundant, wino-like neutralinos, CTA can provide a good way of indirectly detecting them in both types of searches.

In \reffig{fig:3b} we show the results that correspond to a scenario with the neutralino being the only DM particle and having
its production in the early Universe supplemented by, \textit{e.g.}, some non-thermal contribution.
Notably, this allows one to consider neutralino DM with significantly larger values of the annihilation cross section and,
therefore, much better prospects for discovery in future indirect searches. In particular, in this scenario CTA could easily discover higgsino-like neutralino DM with the mass of order a few hundred GeV in both diffuse photon and monochromatic-line searches. As can be seen in the plot, the Fermi-LAT limits\cite{Fermi-LAT:2016uux} bite
into the low mass region of the parameter space, where IACTs lose sensitivity. This is illustrated in   \reffig{fig:3b} by a dashed line for fixed annihilation final state into a $b\bar{b}$ pair, which well represents the position of the exclusion bound we would obtain when imposing Fermi-LAT as a constraint in the likelihood.
This scenario is also independently constrained by DD searches of DM, which are taken into account in our scanning procedure.

%%%%%%%%%%%%%%%%%%%%%%%%%%%%%%%%%%%%%%%%%%%%%%%%%%%%%%%%%%%%%%%%%%%%%%
\subsection{Study of benchmark points\label{sec:benchmarks}}

In the previous section we have computed the H.E.S.S. limits and CTA sensitivity in the p9MSSM
by combining the bounds shown in \reffig{fig:linesa} weighted by the branching fractions to the appropriate final states.
We have already noted, however, that, in principle, a more robust
procedure should be applied. It would involve summing over all weighted spectra of annihilation final states and then using up-to-date instrument response functions and background estimates to obtain the limit as described in detail in \refsec{sec:DMIDgamma}. The full procedure, on the other hand, has the disadvantage of being extremely
time and CPU consuming. In this section, we test the simplified treatment against the more accurate one for some carefully selected representative benchmark scenarios.

For that purpose, we choose 7 benchmark points with different neutralino properties and diverse annihilation final states.
The physical properties of these points are summarized in \reftable{tab:br}.
Among these points, BM5$-$BM7 fail to provide the thermally produced relic abundance consistent with $\Omega_\chi h^2\approx 0.12$
in the standard freeze-out scenario, but could do this, \textit{e.g.}, in modifed cosmological scenarios.
The last two rows of the table show the difference between the 95\%~C.L. CTA projected sensitivities computed with the simplified and full procedures.

\begin{figure}[ht]
\centering
\includegraphics[width=0.8\textwidth]{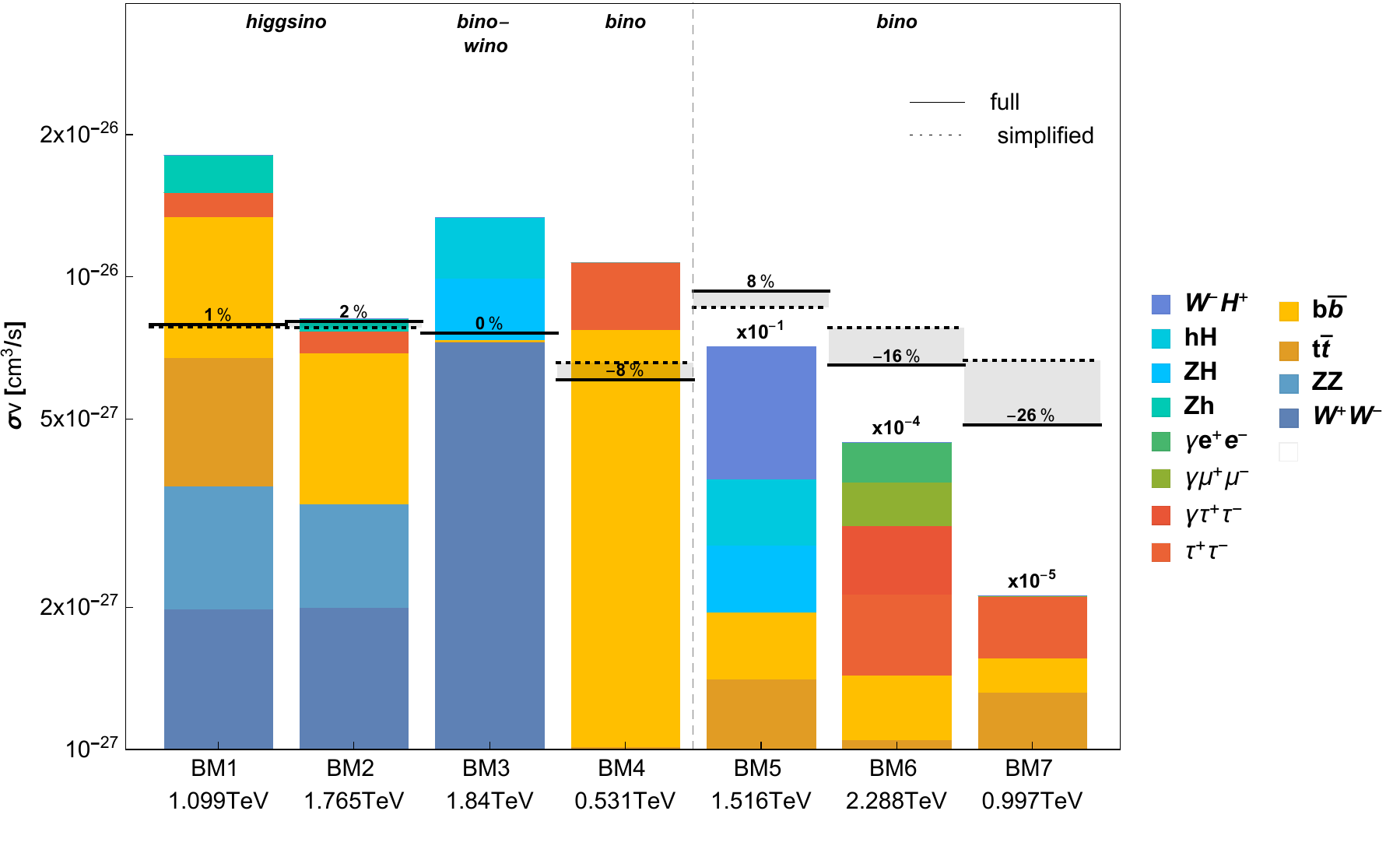}
\caption{\footnotesize Comparison between $\sigma v_{0}^{95\%\,\textrm{C.L. (full calc.)}}$ and $\sigma v_{0}^{95\%\,\textrm{C.L. (simplified)}}$ for the 7 selected representative benchmark points given in \reftable{tab:br}.
The color code tracks the branching fractions for dominant present-day annihilation channels,
while the height of the column gives the theoretical value of the cross section $\sigma v_0$.
The relative difference between the full and simplified procedure is highlighted in percentage.
Benchmark points to the right of the vertical dashed line do not yield the correct relic abundance
and feature annihilation cross section much below the projected limits. They are characterized by a large number of differentiated  annihilation channels and, in particular, include large fraction to $\tau^+\tau^-$. These are the spectra producing the maximal
difference between the two computational methods. It follows as a consequence
that in the physically relevant region analyzed in \refsec{sec:results} the simplified procedure gives a very good approximation of the full calculation.}
\label{fig:BP}
\end{figure}

The dependence of this difference on final states and specific branching ratios is shown in \reffig{fig:BP}.
We find good agreement, better than 10\%, for typical points corresponding to higgsino-like, wino-like, mixed bino-wino and some
pure bino-like neutralinos. The biggest discrepancy (up to $25\%$) occurs for BM6 and BM7 which are bino-like neutralinos that
annihilate primarily to leptons (note the different shape of the limit for the $\tau^+\tau^-$ annihilation final state in \reffig{fig:linesa}) but that also exhibit a significant branching fraction into hadronic final states. Such points are not found in \reffig{fig:pMSSM1}$-$\reffig{fig:3b} as their thermal relic density would overclose the Universe.
Moreover, their $\sigma v_0$ is  orders of magnitude below the CTA sensitivity, hence they would be irrelevant for determining CTA prospects of detecting neutralino DM within the p9MSSM,
even if their relic density was altered in the desired way by assuming a modified cosmological history.

However, it is interesting to note that this discrepancy is not due to the low statistics of the signal coming from neutralinos with
a small value of the annihilation cross section.
It actually persists if one multiplies $\sigma v_0$ by an appropriate factor that brings $\sigma v_0$ close to the projected CTA sensitivity reach. Therefore, it could potentially also affect some analyses performed for other models of new physics in which a particle DM candidate features mixed leptonic-hadronic final annihilation states, and could lead to a sizable discrepancy between the true reach of indirect detection experiments and the one determined by the simplified approach (or similar).

\begin{table}[h]
\centering
\resizebox{\textwidth}{!}{\begin{tabular}{|c|c|c|c|c|c|c|c|}
\hline
Benchmark   &  BM1   &  BM2  &  BM3  &  BM4  &  BM5  &  BM6   &  BM7  \\
points  &  &  &  &  &  &  &  \\
 \hline
$m_{\chi}$  &   1099  &  1765   &  1840  &  531  &  1516   &    2288   &  997 \\
$\rm [GeV]$  &  &  &  &  &  &  &  \\
 \hline
Branching   &  $W^+W^-$ 0.23  &  $b\overline{b}$ 0.35  &  $W^+W^-$ 0.64    &  $b\overline{b}$  0.85  &  $t\overline{t}$ 0.17 &  $\tau^+\tau^-$ 0.26   &  $\tau^+\tau^-$ 0.39  \\
fractions  &  $b\overline{b}$ 0.23  &   $W^+W^-$ 0.29 &   $hA$ 0.14  &  $\tau^+\tau^-$ 0.14  &  $b\overline{b}$  0.16 &  $\gamma\tau^+\tau^-$ 0.22   &  $t\overline{t}$ 0.37 \\
 &  $t\overline{t}$  0.21 &  ZZ 0.24 &   ZH  0.14  &  $t\overline{t}$ 0.01   &  $hA$  0.16 &  $b\overline{b}$ 0.21   &  $b\overline{b}$  0.22 \\
 &  ZZ 0.20 &  $\tau^+\tau^-$ 0.05 &   $\gamma W^+W^-$ 0.08  &    &  ZH 0.16 &  $\gamma\mu^+\mu^-$  0.14   &  $\gamma\tau^+\tau^-$ 0.01 \\
 &  Zh 0.06  &  $\gamma W^+W^-$ 0.04 &     &   &  $W^+H^-$ 0.16 &  $\gamma e^+e^-$ 0.13   &    \\
 &  $\tau^+\tau^-$ 0.04 &  Zh  0.03  &     &   &  $W^-H^+$ 0.157 &  $t\overline{t}$ 0.03   &   \\
  \hline
$\Omega_{\chi} h^2$    &    0.10   &  0.16  &   0.13   &  0.13   &   -   &  -  &   -   \\
 \hline
Main    &  \charone\   &  $\chi\chi \to \text{SM}$  &   \charone, $\chi_2$   &   $\chi\chi \to \text{SM}$  &  \charone, $\chi_2$   &   $\chi\chi \to \text{SM}$  &  slepton  \\
mechanism  &    coann.   &  $A$-funnel  &   coann.   &  t-channel  &   coann.   &  t-channel  &   coann.  \\
\hline
$\sigma v_0$   &  $1.98\cdot 10^{-26}$  &  $8.18\cdot 10^{-27}$  &  $1.08\cdot 10^{-26}$  &  $1.12\cdot 10^{-26}$  &  $7.63\cdot 10^{-28}$  &  $4.54\cdot 10^{-31}$    &  $2.14\cdot 10^{-32}$ \\
$\rm [cm^3s^{-1}]$  &  &  &   &   &  &  & \\
 \hline
$\sigma v_{0}^{95\%\,\textrm{C.L. (simplified)}}$  &  $7.64\cdot 10^{-27}$   &  $7.82\cdot 10^{-27}$  &   $6.59\cdot 10^{-27}$   &  $7.29\cdot 10^{-27}$  &  $8.84\cdot 10^{-27}$  &  $7.82\cdot 10^{-27}$   &  $6.65\cdot 10^{-27}$ \\
$\rm [cm^3s^{-1}]$  &  &  &   &   &  &  & \\
 \hline
$\sigma v_{0}^{95\%\,\textrm{C.L. (full calc.)}}$   &  $7.92\cdot10^{-27}$   &  $8.05\cdot10^{-27}$  &  $6.06\cdot 10^{-27}$  &  $7.47\cdot 10^{-27}$  &  $9.34\cdot10^{-27}$  &  $6.50\cdot10^{-27}$   &  $4.87\cdot10^{-27}$ \\
$\rm [cm^3s^{-1}]$  &  &  &   &  &  &  & \\
 \hline
\end{tabular}}
\caption{\footnotesize Selected benchmark points characterized by different properties.
The main annihilation mechanism at freeze-out corresponds to the final state with the largest branching ratio. The theoretical value of the cross section $\sigma v_0$ is given as well as the relic density $\Omega_{\chi} h^2$.
CTA sensitivity is reported for the simplified and full scheme computation.
The sensitivity is expressed as 95\%~C.L. upper limits. The LLRTS value is derived according to Eq.~(\ref{TS}) for the given
$\sigma v_0$ in the two computation schemes.}
\label{tab:br}
\end{table}

%%%%%%%%%%%%%%%%%%%%%%%%%%%%%%%%%%%%%%%%%%%%%%%%%%%%%%%%%%%%%%%%%%%%%%
\section{Conclusions} \label{sec:Conclusions}

In this work we performed an updated and improved study of the reach
of CTA in testing neutralino DM in minimal supersymmetric
scenarios. The results were compared with the most recent bounds on $\sigma v_0$, as a function of DM mass, obtained by H.E.S.S.
We conducted the analysis in the framework of the 9-parameter MSSM, or p9MSSM.
We included the most recent constraints from DM direct detection searches, flavor physics, and Higgs searches, and constructed
a state-of-the-art likelihood ratio test statistic approach to
analyze the CTA sensitivity. The direct constraints on sparticle masses from the LHC are also included, although they are known to be of very limited impact for the parameter space leading to TeV-scale DM.
Furthermore, on the theoretical side we refined the calculations of DM
relic abundance and present-day annihilation cross section by taking
into account the Sommerfeld enhancement effect for a completely
generic mixed neutralino and its co-annihilation partners. In
particular, for the first time sfermion co-annihilations were
considered with  Sommerfeld effect included in a scanning framework.

Having all these improvements implemented, we performed numerical scans of the p9MSSM parameter space focusing on a TeV scale neutralino DM. We find that, assuming the Einasto profile of DM halo in the Milky Way, H.E.S.S. has been able to nearly reach the so-called thermal WIMP value, while CTA will go below it by providing a further improvement of at least an order of magnitude. The results show that both H.E.S.S. and
CTA are sensitive to several cases for which direct detection cross
section will be below the so-called neutrino floor, with
H.E.S.S. being sensitive to most of the wino region, while CTA also
covering a large fraction of the 1\tev\ higgsino region. We
additionally show the extent to which the CTA sensitivity will be further improved in the monochromatic photon search mode for both single-component and underabundant DM.

While we focused on the Einasto profile when presenting the results
for the p9MSSM, we also studied two other DM profiles, namely the
standard NFW profile and the version of the Einasto profile with a
core with conservative radius $r_c=3$~kpc, for which we presented the most
up-to-date CTA sensitivities in searches relevant for a number of
fixed annihilation final states. These can be easily combined to
derive actual results for any model of new physics predicting heavy
WIMP DM. In particular, when applied to the p9MSSM, the aforementioned
Cored Einasto profile leads to substantially weaker current bounds and
future sensitivity reaches. In this case, the H.E.S.S. limits do not
completely exclude the region of the parameter space with wino-like
neutralino DM. Instead, CTA will be able to fully probe this important
scenario.

%%%%%%%%%%%%%%%%%%%%%%%%%%%%%%%%%%%%%%%%%%%%%%%%%%%%%%%%%%%%%%%%%%%%%%
\acknowledgments
We would like to thank Alexander Pukhov for providing a refined \texttt{micrOMEGAs} module that calculates annihilation cross section to monochromatic photons. We would like to thank Martin Vollmann for useful discussions. This research has made use of the CTA instrument response functions provided by the CTA Consortium and Observatory, see \url{http://www.cta-observatory.org/science/cta-performance/} (version prod3b-v1) for more details. K.J., L.Rosz. and S.T. are supported by the National Science Centre (NCN) research grant No. 2015/18/A/ST2/00748. L.Rosz. is also supported by the National Science Centre research grant No. 2016/22/M/ST9/00583 and by ``AstroCeNT: Particle Astrophysics Science and Technology Centre” project that is carried out within the International Research Agendas programme of the Foundation for Polish Science co-financed by the European Union under the European Regional Development Fund. E.M.S. is supported in part by the National Science Centre (Poland) under the research Grant No.~2017/26/D/ST2/00490. S.T. is supported in part by the
Lancaster-Manchester-Sheffield Consortium for Fundamental Physics under STFC grant:
ST/P000800/1. The use of the CIS computer cluster at the National Centre for Nuclear Research in Warsaw is gratefully acknowledged.

%%%%%%%%%%%%%%%%%%%%%%%%%%%%%%%%%%%%%%%%%%%%%%%%%%%%%%%%%%%%%%%%%%%%%%
\bibliographystyle{JHEP}
%\bibliography{CTA_MSSM_2018}

\providecommand{\href}[2]{#2}\begingroup\raggedright\endgroup

\end{document}